\documentclass[12pt]{article}
\usepackage{cite,graphicx,amsmath,amssymb}
\usepackage{graphicx}
\usepackage{multicol}
\usepackage{xcolor}
\usepackage{multirow}
\usepackage{slashed}

\newcommand{\be}{\begin{equation}}
\newcommand{\ee}{\end{equation}}
\newcommand{\bea}{\begin{eqnarray}}
\newcommand{\eea}{\end{eqnarray}}
\begin{document}

\thispagestyle{empty}
\vspace*{.2cm}
\noindent
\hfill 18 May 2012

\vspace*{2.0cm}

\begin{center}
{\Large\bf NMSSM with Lopsided Gauge Mediation}
\\[2.5cm]

{\large Ivan Donkin, Alexander K. Knochel
\\[6mm]}

{\it
Institut f\"ur Theoretische Physik, Universit\"at Heidelberg, 
Philosophenweg 19,\\ D-69120 Heidelberg, Germany\\[3mm]

{\small\tt (\,donkin@thphys.uni-heidelberg.de} %{\small ,} 
{\small\tt \,a.k.knochel@thphys.uni-heidelberg.de} 
\small\tt ) }
\\[2.0cm]

{\bf Abstract}
\end{center} 

\noindent
We study a gauge mediated supersymmetry breaking version of the NMSSM in which the soft $m_{H_u}^2$ and $m_{H_d}^2$ masses receive extra contributions due to 
the presence of direct couplings between the Higgs and the messenger sector. We are motivated by the 
well-known result that minimal gauge mediation is phenomenologically incompatible with the NMSSM due to 
the small value of the induced effective $\mu$ term. The model considered in the present paper solves 
the aforementioned problem through a modified RG running of the singlet soft mass $m_N^2$. This effect, which 
is induced by the dominant $m_{H_d}^2$ term in the one-loop $\beta$-function of $\,m_N^2\,$, shifts the 
singlet soft mass towards large negative values at the electroweak scale. That is sufficient to ensure a large VEV 
for the scalar component of the singlet which in turn translates into a sizeable effective $\mu$ term. We also 
describe a mechanism for generating large soft trilinear terms at the messenger scale. This allows us to make the 
mass of the lightest Higgs boson compatible with the current LHC bound without relying on exceedingly heavy stops.

\newpage

\section{Introduction}

Gauge mediation constitutes a very predictive framework which describes the transmission of SUSY 
breaking effects from the hidden to the observable sector \cite{Giudice:1998bp, Gaume:1982cw, Dine:1993yw}. 
Among other things it guarantees flavor universality of the soft sfermion masses at the messenger scale and 
therefore suppresses the potentially dangerous flavor-changing neutral currents \cite{Ellis:1981ts}.

However, in its minimal form gauge mediation has nothing to say about the origin of the $\mu$ and $B_{\mu}$ 
terms. The problem arises as follows: Recall that in MGM models 
SUSY breaking effects are transmitted from the hidden to the observable sector only through gauge interactions.
In the absence of direct couplings between the messenger and the Higgs sector, Peccei-Quinn is an exact symmetry 
of the fundamental theory. The Higgs bilinear term $\,\mu H_u H_d\,$, on the other hand, breaks this symmetry 
explicitly \cite{Peccei:1977hh, Peccei:1977ur}. Therefore this operator cannot be present in the low-energy 
effective action which arises after integrating out the heavy messenger fields at one loop. In this sense the MGM 
scenario is not considered satisfactory unless one specifies the extra mechanism responsible for generating $\mu$ 
and $B_{\mu}$.

There are various proposals in the literature for generating a Higgs bilinear term of the correct order of 
magnitude. In the context of the so called Giudice-Masiero solution, which was initially formulated in models 
with gravity mediated supersymmetry breaking (see \cite{Guidice:1988gm}), a tree-level $\mu$-term is forbidden from the 
original superpotential and is induced by the following higher-dimensional effective operators in the K\"ahler potential:
\begin{equation}\label{GiuMas}
K \supset \frac{c_1}{M} \, \int d^4\theta H_u H_d X^{\dagger} \, + \, \frac{c_2}{M^2}\,\int d^4\theta
H_u H_d X X^{\dagger} \,\,+\,\,\,\, \mbox{h.c.}
\end{equation}
The parametrization in eq.(\ref{GiuMas}) applies both to models with gauge and  gravity
mediated supersymmetry breaking. In the GMSB context $X$ is a spurion superfield which couples to the hidden sector, 
$M$ is the messenger scale and the prefactors $c_1$ and $c_2$ stand for the product of coupling constants and 
possible loop factors. The interactions in eq.(\ref{GiuMas}) can be generated radiatively after introducing 
direct couplings between the Higgs and the messenger sector and integrating out the heavy messenger superfields
at one loop. Substituting $X$ by its $F$-term VEV, we obtain a $\,\mu = c_1 \Lambda\,$ term from the first 
operator in eq.(\ref{GiuMas}) and a $\,B_{\mu} = c_2 \Lambda^2\,$ term from the second operator (with $\Lambda = F_X/M$ 
being the effective scale of SUSY breaking). In theories with gravity mediated supersymmetry breaking one can assume that 
the prefactors $c_1$ and $c_2$ are of order one, $c_1 \sim c_2 \sim \mathcal{O}(1)$, which leads to the celebrated Guidice-Masiero 
solution $B_{\mu} \sim \mu^2$. In the gauge mediated supersymmetry breaking setup, on the other hand, $\mu$ and $B_{\mu}$ are generated 
at one loop and one anticipates that $\,c_1 \sim c_2 \sim 1/16\pi^2$. Therefore the second power $\,\mu^2\,$ is loop suppressed with 
respect to $B_{\mu}\,$, i.e. $B_{\mu} \sim 16 \pi^2 \, \mu^2\,$. This leads to the relation
\begin{equation}\label{gmsbproblem}
B_{\mu} \gg \mu^2
\end{equation}
which is incompatible with electroweak symmetry breaking due to the unacceptably large $B_{\mu}$. 
For this reason eq.(\ref{gmsbproblem}) is usually referred to as the $\mu/B_{\mu}$ problem of gauge-mediated 
supersymmetry breaking \cite{Dvali:1996cu}.

Alternative solutions, which avoid eq.(\ref{gmsbproblem}) and ensure that $B_{\mu} \sim \mu^2$, were formulated e.g. 
in the context of the so called dynamical relaxation mechanism \cite{Dvali:1996cu}. More recently, the authors of \cite{Csaki:2008sr} 
have pointed out that successful EWSB can be achieved even if the problematic relation $B_{\mu} \gg  \mu^2$ is kept 
intact provided that one assumes a non-trivial hierarchy between the mass terms in the Higgs sector at the electroweak scale:
\bea\label{hierarchy}
\mu^2 \sim m_{H_u}^2 \ll B_{\mu} \ll m_{H_d}^2 \,\,.
\eea
As was argued in \cite{Csaki:2008sr} the pattern in eq.(\ref{hierarchy}) can be obtained by considering 
models with gauge mediated supersymmetry breaking in which the two Higgs doublets couple directly to messenger 
superfields (or more generally to superfields in the SUSY breaking sector).

In the present paper we analyze eq.(\ref{hierarchy}) in the context of the Next to Minimal Supersymmetric Standard Model 
(NMSSM) \cite{Ellwanger:2009dp}. We are motivated by the well-known result that the NMSSM is incompatible with the minimal form of 
gauge mediated supersymmetry breaking \cite{deGouvea:1997cx} due to the small value of the generated $\,\mu\,$ term. We will argue 
that a phenomenologically viable model can be obtained by a slight deformation of the hierarchy between the soft mass terms in the 
Higgs sector. To be exact, we consider an NMSSM in which the $m_{H_d}^2$ term dominates all other soft masses, in particular we require that $m_{H_d}^2 \gg m_{H_u}^2$. 
We describe an embedding of our scenario into a GMSB setup with direct Higgs-messenger mixing and refer to this model as lopsided NMSSM in accord with 
the terminology introduced in \cite{DeSimone:2011va}. The crucial difference between our approach and the more conventional lopsided MSSM  is the origin of 
the effective $\mu$ term. In the minimal setup considered in \cite{DeSimone:2011va} $\,\mu$ arises solely from the tree-level Higgs-messenger couplings. 
In our setting there is a large extra contribution coming from the NMSSM operator $\lambda N H_u H_d$. As a result the $\mu$ term in the 
lopsided NMSSM is far less constrained than its MSSM counterpart.

The paper is organized as follows: In section 2 we introduce the lopsided GMSB scenario following closely the approach in
\cite{Csaki:2008sr}. Using this line of reasoning we define a lopsided NMSSM which is the main object 
of study in the remainder of the paper. In section 3 we give a detailed description of the field content and superpotential 
couplings of our theory, including a full specification of the hidden and messenger sectors. In section 4 we show that the 
effective $\mu$ term is comprised of two pieces -- one arising from the Higgs-messenger mixing and another associated with the
NMSSM part of the superpotential. We describe in detail the mechanism which triggers a large vacuum expectation value for the 
NMSSM singlet and therefore generates the latter piece of the Higgsino mass parameter. In section 5 we calculate the low-energy spectrum for
several points in parameter space. Crucially, we describe a new way for generating large soft trilinear terms at the messenger scale.
As pointed out already in \cite{Draper:2011aa} this is chiefly important for raising the Higgs mass above the current LHC bound. 
We also find that, for large regions of parameter space, the negative tree-level contribution to $m_{h^0}$ from the mixing with 
the singlet is unacceptably large and discuss a possible mechanism which can suppress this effect. In section 6 we briefly discuss 
the collider phenomenology of the model.

\section{Defining the lopsided NMSSM}

In order to set the stage for our subsequent discussion and fix the notation we briefly recall the basics of lopsided gauge 
mediation. Here we follow the approach in \cite{Csaki:2008sr}, i.e. we introduce the lopsided MSSM as a possible 
solution to the $\mu/B_{\mu}$ problem. Let us analyze eq.(\ref{hierarchy}) in more detail. To this end we shall look at the
tree-level minimization conditions in the Higgs sector:
\bea\label{higgs1}
\frac{m_Z^2}{2} = - |\mu|^2 - \frac{m^2_{H_u} \tan^2 \beta - m^2_{H_d}}{\tan^2 \beta -1}
\eea
\bea\label{higgs2}
\sin 2\beta = \frac{2 B_{\mu}}{2|\mu|^2 + m_{H_u}^2 + m_{H_d}^2}
\eea
One can argue that the above hierarchy leads to a fully natural electroweak symmetry breaking. Indeed, taking into consideration
that $\sin 2\beta \approx 2 B_{\mu} / m_{H_d}^2$, we can rewrite (\ref{higgs2}) as a quadratic equation for $\tan \beta$ 
which leads to
\bea
\tan \beta \approx \frac{m_{H_d}^2}{B_{\mu}} \gg 1\,\label{tan}.
\eea
This shows that the proposed solution is operational only in the large $\tan\beta$ regime. From eq.(\ref{higgs1}) we then obtain:
\bea
\frac{m_Z^2}{2} = - |\mu|^2 - m^2_{H_u}  - \frac{m^2_{H_d}}{\tan^2 \beta} \label{min}
\eea
Note that one can make all terms on the right hand side roughly of the same order of magnitude, while still keeping the hierarchy 
in eq.(\ref{hierarchy}) intact. This is most easily seen by considering the 
parametrization from \cite{Csaki:2008sr} 
\bea
\mu = \epsilon \Lambda_{\rm Higgs} \quad\quad B_{\mu} = \epsilon \Lambda^2_{\rm Higgs} \quad\quad
m^2_{H_u} = \epsilon^2 \Lambda^2_{\rm Higgs} \quad\quad m^2_{H_d} = \Lambda^2_{\rm Higgs} 
\eea
where $\,\epsilon \ll 1\,$ is some small number and $\,\Lambda_{\rm Higgs}\,$ denotes the effective scale of SUSY breaking in the 
Higgs sector. Noting that $\tan \beta = 1/\epsilon$ we deduce that all terms on the right hand side of eq.(\ref{min}) are of the 
order $\epsilon^2 \Lambda_{\rm Higgs}^2$. Thus we expect that in the absence of large radiative correction electroweak symmetry breaking 
should occur naturally, without the need for significant fine-tuning.

The authors of \cite{Csaki:2008sr} provide a possible mechanism for generating the pattern in eq.(\ref{hierarchy}). 
Specifically, we can envisage a non-minimal gauge-mediated scenario with direct couplings between the Higgs and the messenger sector 
of the form
\bea\label{bilinears}
W \supset  \lambda_u H_u \Phi_1 \bar{\Phi}_2 + \lambda_d H_d \bar{\Phi}_1 \Phi_2  \, + \, {\rm h.c} \,\,\,.
\eea
Here $\,\Phi_1 \bar{\Phi}_2\,$ and $\,\bar{\Phi}_1 \Phi_2\,$ are bilinears of messenger fields. After integrating out the the heavy fields represented by
the vector-like pairs $\,\Phi_1$, $\bar{\Phi}_1\,$ and $\,\Phi_2$, $\bar{\Phi}_2\,$ we obtain the following contributions to the mass parameters in the Higgs 
sector
\bea\label{additional}
\mu \approx \lambda_u \lambda_d \, \frac{N_{\rm Higgs}}{16\pi^2} \,\Lambda_{\rm Higgs} \,, 
\quad B_{\mu} \approx \lambda_u \lambda_d \frac{N_{\rm Higgs}}{16\pi^2} \, \Lambda_{\rm Higgs}^2\,, 
\quad m_{H_{u,d}}^2 \approx \lambda_{u,d}^2 \frac{N_{\rm Higgs}}{16\pi^2}\,\Lambda_{\rm Higgs}^2
\eea
where $\,N_{\rm Higgs}\,$ parametrizes the number of messenger fields coupled to the Higgs sector (i.e. $\,N_{\rm Higgs} 
= n_{5+\overline{5}} +   3 n_{10 +\overline{10}}\,$ is the ``messenger index'' in the Higgs sector). Observe that $\,\Lambda_{\rm Higgs}\,$ 
could but does not have to coincide with the SUSY breaking scale $\Lambda$ from the MGM sector. The same statement applies to $N_{\rm Higgs}$.
Imposing the hierarchy $\lambda_d \gg \lambda_u$ and introducing the parameter $\,\epsilon = \frac{N_{\rm Higgs}}{16 \pi^2}\,\frac{\lambda_d}{\lambda_u}\,$ 
leads to the correct low-energy pattern from eq.(\ref{hierarchy}). In \cite{DeSimone:2011va} the class of models specified by eqs.(\ref{hierarchy}) 
and (\ref{bilinears}) was labeled lopsided gauge mediation.

In the present section we introduce the lopsided NMSSM as an effective  theory, i.e. we do not care how supersymmetry is broken or transmitted
to the observable sector. To begin the description of our model let us fix the superpotential in the Higgs sector:
\bea
W \supset (\mu + \lambda N) H_u H_d  + \frac{\kappa}{3}\,N^3 \,\,.\label{lopsidedpotential}
\eea
Recall that the operator $\,\lambda N H_u H_d\,$  is invariant under a Peccei-Quinn symmetry 
which acts on the $\,N, H_u, H_d\,$ superfields according to 
\begin{equation}
H_u \to H_u \, e^{-i\phi} \quad\quad H_d \to H_d \, e^{-i\phi} \quad\quad N \to N e^{i 2\phi} \,\,.
\end{equation}
This symmetry is broken at tree-level in a twofold manner -- both by the cubic $\,\frac{\kappa}{3}\, N^3$ 
operator and the bilinear mass term for the $H_u$, $H_d$ Higgs doublets. As a result the potentially dangerous
Peccei-Quinn axion in the Higgs sector is automatically avoided. There is an additional subtlety -- the $\,\frac{\kappa}{3}
\,N^3\,$ term is invariant under a residual $\mathbb{Z}_3$ symmetry (the action of the $\mathbb{Z}_3$ group rotates all 
three superfields $N$, $H_u$ and $H_d$ by the same phase $e^{2\pi i/3}$). In the standard $\mathbb{Z}_3$-invariant NMSSM 
this symmetry is responsible for the appearance of domain walls  \cite{Abel:1995wk, Derendinger1984307, Ellwanger:2009dp}. 
Note, however, that in the setup we are considering the discrete $\mathbb{Z}_3$ is not an actual symmetry of the Lagrangian -- it is broken 
at tree-level by the presence of a $\mu$-term. Therefore the domain wall problem is not present in our version of the lopsided NMSSM scenario.
\footnote{Clearly one can consider a lopsided NMSSM without a tree-level $\mu$ term. In this case the domain wall problem has to be solved through an alternative mechanism.} 

The soft supersymmetry breaking terms in the Higgs sector read
\bea\label{soft}
-\mathcal{L}_{\rm soft} \,=\,&&\, m_N^2 |N|^2 \, + \,m_{H_u}^2 |H_u|^2 \,+ \,m_{H_d}^2 |H_d|^2 \,+ (\,\lambda A_{\lambda} 
N H_u H_d \, + \, \frac{1}{3}\,\kappa A_{\kappa} N^3 \, + \, {\rm h.c.}\,) \, + \nonumber\\
&&+\,\,\,(\,B_{\mu} H_u H_d \,+ \, {\rm h.c.} \,)
\eea 

Imposing $\,m_{H_d}^2 \gg m_{H_u}^2\,$ \textit{at the electroweak scale} completes the description of the effective Lagrangian. From a low-energy point of view
the precise mechanism which produces the extra contributions to $\,m_{H_d}^2\,$ and $\,m_{H_u}^2\,$ to ensure the non-trivial hierarchy between the soft mass 
terms in the Higgs sector is unimportant. The same statement applies to the origin of the $\mu$ and $B_{\mu}$ terms in eqs.(\ref{lopsidedpotential}) 
and (\ref{soft}). We can simply assume that they are present in the Lagrangian and then extract their values from the tree-level minimization conditions in the 
Higgs sector. In the next section we will construct a high-energy model which produces the lopsided NMSSM as its low-energy limit.

\section{High-energy completion in the context of gauge mediation}

In this section we will embed the lopsided NMSSM into a gauge-mediated scenario. Our guiding principle is very simple -- the model we
are considering should, upon integrating out the messenger superfields at one loop, produce the lopsided NMSSM as its low-energy limit. 
The main building block of our construction is the following superpotential considered in \cite{DeSimone:2011va}:
\bea\label{lpotential}
W &\supset& \, X \Phi \bar{\Phi} \, + \,\lambda N H_d H_u \,+ \,\frac{\kappa}{3} \,N^3 \, + \\
&+&\,\lambda_u H_u D \, T \, + \, \lambda_d H_d \bar{D} \,\bar{T} \, +  X_D D \bar{D}  \, + \, \frac{X_T}{2} \, ( \,T \,\, \bar{T}\,) 
\left( \begin{array} {cc}
a_T          & a_{T\bar{T}} \\
a_{T\bar{T}} & a_{\bar{T}} 
\end{array} \right)   \left( \begin{array} {c}
                                      T \\
                                      \bar{T}
                                      \end{array} \right)   \nonumber
\eea

The operator $X \Phi \bar{\Phi}$ comprises the universal part of the messenger sector which is present in any model of gauge mediation. This part 
consists of messenger superfields $\Phi, \bar{\Phi}$ filling complete vector-like representation of the GUT group $SU(5)$. The 
masses and splittings between the $\Phi, \bar{\Phi}$ are controlled by a spurion superfield $X$ which acquires VEVs in its scalar 
and $F$-term components due to some dynamics in the hidden sector:
\bea
X = M + F \,\theta^2 \,\,.
\eea
After integrating out $\Phi, \bar{\Phi}$ at the messenger scale $M$, gaugino masses are generated at one and sfermion masses at two loop
\bea
M_i = \frac{\alpha_i}{4\pi} \, n \, \Lambda \quad\quad\quad m_{\tilde{f}}^2 = 2 \, n \, \Lambda^2 \, \left[\,\sum_{i=1}^{3} C_{i}^{\tilde{f}} 
\left( \frac{\alpha_i}{4\pi} \right)^2 \, \right] \,\,. \label{gm}
\eea
Here $\Lambda = F/M$ is the effective SUSY breaking scale and $n$ stands for the messenger index. The $C_{i}^{\tilde{f}}$ denotes the quadratic Casimir 
operator for the respective representation of the sfermion superfield $\,\tilde{f}\,$ with respect to the SM gauge group $G_i\,$, where $\,G_1 = U(1)$, 
$G_2 = SU(2)\,$ and $\,G_3 = SU(3)$.

The model contains an NMSSM-like sector with an extra singlet $N$ which couples to the Higgs doublets through the usual NMSSM operator 
$\,\lambda N H_u H_d\,$. To break the PQ symmetry in the Higgs sector we introduce the cubic term $\frac{k}{3}  N^3$ to the superpotential. At this stage
the soft Higgs masses $m_{H_d}^2$ and $m_{H_u}^2$ are equal at the messenger scale. The extra contributions to $m_{H_d}^2$ and $m_{H_u}^2$ which fix the hierarchy 
$m_{H_d}^2 \gg m_{H_u}^2$ come from direct couplings between the Higgs and the messenger sector  $\,\lambda_u H_u \,D T \, + \, \lambda_d H_d \,\bar{D} \bar{T}$. 
Here $\,D, \bar{D}\,$ are messenger superfields in the fundamental representation of $SU(2)$ with opposite hypercharges: 
$D = (\mathbf{1,2})_{1}, \,\bar{D} = (\mathbf{1,2})_{-1}$ (where the subscript stands for the $U(1)_Y$ charge). In addition we 
have two total singlets $T$ and $\bar{T}$ with quantum numbers $T=(\mathbf{1,1})_0\,$ and $\,\bar{T}=(\mathbf{1,1})_0$. In order to preserve gauge
coupling unification we assume that the $D, \bar{D}$ pair is lying in one $\,\mathbf{5}{\oplus}\mathbf{\overline{5}}\,$ copy of $SU(5)$. As usual 
we make the assumption that the two spurions acquire VEVs in their scalar and $F$-term components 
\bea
X_{D} = M_D (\,1 + \Lambda_D \theta^2 ) \quad\quad\quad\quad   X_T = M_T (\,1 + \Lambda_T \theta^2) \,\,.
\eea
The messenger sector of the above superpotential (the second line in eq.(\ref{lpotential})) exhibits a discrete $\mathbb{Z}_2$ symmetry in the limit $a_{T\bar{T}} = 0$. The action of the $\mathbb{Z}_2$ group is 
given by:
\bea
H_u \to - H_u \quad\quad\quad T \to - T
\eea
Note that this symmetry forbids a $\mu$ (and a corresponding $B_{\mu}$ term) in the effective low-energy action. This means that we can control
the magnitude of $\mu$ and $B_{\mu}$ by suitably adjusting the value of the $a_{T\bar{T}}$ coupling. Following \cite{DeSimone:2011va} we can
diagonalize the $T,\bar{T}$ mass matrix with a suitably chosen rotation by an angle $\Theta$. In the following we shall denote the eigenvalues of 
the mass matrix by $p$ and $\,\xi\, p$ where $\,\xi\,$ is the proportionality factor. The contribution to the mass parameters in the Higgs sector 
coming from the superpotential (\ref{lpotential}) have been calculated in \cite{DeSimone:2011va}. Explicitly they read 
\bea
&&m_{H_{u}}^2 \, = \, \frac{\lambda_{u}^2}{16\pi^2} \, \Lambda_D^2 \, \left( \,\cos^2\Theta \, P(x,y) + 
\sin^2\Theta \, P(\xi x,y)\,\right) = a_u \, \frac{\lambda_{u}^2}{16\pi^2} \, \Lambda_D^2 \label{firstcoeff}\\
&&m_{H_{d}}^2 \, = \, \frac{\lambda_{d}^2}{16\pi^2} \, \Lambda_D^2 \, \left( \,\sin^2\Theta \, P(x,y) + 
\cos^2\Theta \, P(\xi x,y)\,\right) = a_d \,\frac{\lambda_{d}^2}{16\pi^2} \, \Lambda_D^2 \\
&&\mu \, = \, \frac{\lambda_u \lambda_d}{16\pi^2} \, \Lambda_D \,\sin\Theta\,\cos\Theta \,\left( \,-\,Q(x,y)
+ Q(\xi x,y)\,\right) = a_{\mu}\, \frac{\lambda_u \lambda_d}{16\pi^2} \, \Lambda_D \label{muterm} \\
&&B_{\mu} \, = \, \frac{\lambda_u \lambda_d}{16\pi^2}\,\Lambda_D^2 \,\sin\Theta \,\cos\Theta\,\left(\,-\,R(x,y)
+ R(\xi x,y)\,\right) =  a_{B_{\mu}}\,\frac{\lambda_u \lambda_d}{16\pi^2}\,\Lambda_D^2  \\
&&A_{u} \, = \, \frac{\lambda_{u}^2}{16\pi^2}\, \Lambda_D \left(\,\cos^2\Theta \, S(x,y) + \sin^2\Theta\, S(\xi x,y)\,\right)=  
a_{A_{u}}\,\frac{\lambda_u \lambda_d}{16\pi^2}\,\Lambda_D  \\
&&A_{d} \, = \, \frac{\lambda_{d}^2}{16\pi^2}\, \Lambda_D \left(\,\sin^2\Theta \, S(x,y) + \cos^2\Theta\, S(\xi x,y)\,\right)= 
a_{A_{d}}\,\frac{\lambda_u \lambda_d}{16\pi^2}\,\Lambda_D\label{lastcoeff}
\eea
where we used the shorthand notation $x = M_T/M_D$ and $y = \Lambda_T/\Lambda_D$. The coefficient functions $P(x,y)$, $Q(x,y)$, $R(x,y)$ and 
$S(x,y)$ appearing in eqs.(\ref{firstcoeff})--(\ref{lastcoeff}) are listed in Appendix B. We can recover the $\mathbb{Z}_2$-symmetric limit by 
setting $a_{T\bar{T}} = 0$. The crucial feature of the model we are considering is that it allows us to switch off the contributions 
to the soft $m_{H_u}^2$ and $m_{H_d}^2$ masses and/or to the $\mu$ and $B_{\mu}$ terms by an appropriate choice of the $\,\Theta$, 
$\xi\,$ and $\,y$ parameters. For example taking $\,\Theta =0\,$ or $\,\xi=1\,$ leads to $\,\mu = B_{\mu} = 0\,$ whereas fixing $y=1$ 
shuts off the extra contributions to the $m_{H_u}^2$ and $m_{H_d}^2$ soft masses. In both cases non-zero $A_u$ and $A_d$ are generated
at the messenger scale.

With these prerequisites we are now ready to write down the full UV completion of the low-energy effective theory 
presented in section 2. Our idea is to simply take three copies of the second line in eq.(\ref{lpotential}):
\bea\label{fullUV}
W &\supset& \,X\Phi\bar{\Phi} \, + \,\lambda N H_d H_u \,+ 
\,\frac{\kappa}{3} \,N^3 \, + \, \\
&+& \, \lambda_{u,T} H_u D T \, \, \, + \, \lambda_{d,T} H_d \bar{D} \bar{T} \,\,\, + \, X_D D \bar{D}  \, + \, \frac{X_T}{2} \, 
( \,T \,\, \bar{T}\,) \left( \begin{array} {cc}
a_T          & a_{T\bar{T}} \\
a_{T\bar{T}} & a_{\bar{T}} 
\end{array} \right)   \left( \begin{array} {c}
                                      T \\
                                      \bar{T}
                                      \end{array} \right)   \, + \nonumber\\
&+& \lambda_{u,P} H_u F P \, + \, \lambda_{d,P} H_d \bar{F} \bar{P} + X_F F \bar{F}  \, + \, 
\frac{X_P}{2} \, ( \,P \,\, \bar{P}\,) \left( \begin{array} {cc}
a_P          & a_{P\bar{P}} \\
a_{P\bar{P}} & a_{\bar{P}} 
\end{array} \right)   \left( \begin{array} {c}
                                      P \\
                                      \bar{P}
                                      \end{array} \right)  \, + \, \nonumber\\
&+& \lambda_{u,S} H_u E S \, + \, \lambda_{d,S} H_d \bar{E} \bar{S}\, +  \, X_E E \bar{E}  \, + \,\,\, 
\frac{X_S}{2} \, ( \,S \,\, \bar{S}\,) \left( \begin{array} {cc}
a_S          & a_{S\bar{S}} \\
a_{S\bar{S}} & a_{\bar{S}} 
\end{array} \right)   \left( \begin{array} {c}
                                      S \\
                                      \bar{S}
                                      \end{array} \right)  \nonumber
\eea 
The four additional $SU(2)$ doublets $F$, $E$,  $\bar{F}$, $\bar{E}$ have the same quantum numbers as 
$D$ and $\bar{D}$ respectively. The other extra superfields $P, \bar{P}, S, \bar{S}$ are total singlets with 
respect to the SM gauge group. It is implicitly understood that all three pairs $D, \bar{D}$, $F, \bar{F}$ and $E, \bar{E}$
lie in $\,\mathbf{5}{\oplus}\overline{\mathbf{5}}\,$ copies of $SU(5)$. The extra spurions $X_F$ and $X_P$ acquire VEVs in 
their scalar and $F$-term components:
\bea
X_{F} &=& M_F (\,1 + \Lambda_F \theta^2 ) \quad\quad X_{E} = M_E (\,1 + \Lambda_E \theta^2 ) \\
X_P &=& M_P (\,1 + \Lambda_P \theta^2) \quad\quad  X_S = M_S (\,1 + \Lambda_S \theta^2)
\eea
To complete the description of our model we fix
\bea
a_{T\bar{T}} = a_{S\bar{S}} = 0 \quad\quad\quad\quad  \Lambda_P/\Lambda_F = \Lambda_S/\Lambda_E  =1\,\,. 
\eea
With this choice of parameters it is then clear that

1. the second line in eq.(\ref{fullUV}) produces extra contributions only to the $m_{H_u}^2$ and $m_{H_d}^2$ 
masses and the soft trilinear terms $A_{H_u}$, $A_{H_d}\,$,

2. the third line in eq.(\ref{fullUV}) generates non-zero $\mu$, $B_{\mu}$ as well as non-zero $A_{H_u}$, $A_{H_d}$ but does not contribute to either 
$m_{H_u}^2$ or $m_{H_d}^2\,$,

3. the last line in eq.(\ref{fullUV}) does not contribute to any of the $\mu$, $B_{\mu}$, $m_{H_u}^2$, $m_{H_d}^2$ mass terms
in the Higgs sector -- it only generates non-zero soft trilinear terms.

\smallskip
There is an additional subtlety related to the value of the $B_{\mu}$ term in models with lopsided gauge mediation. As pointed out in \cite{DeSimone:2011va} $\,B_{\mu}$ transforms 
under $U(1)_R$ phase rotations while the other scalar mass terms in the Lagrangian do not. In particular we can use the $U(1)_R$ symmetry to forbid a $B_{\mu}$ term altogether. To put it differently -- 
the overall size of $B_{\mu}$ must be a free parameter which should depend on the value of some appropriately chosen coupling in the high-energy model.

What we get after integrating out the messenger superfields from eq.(\ref{fullUV}) is an NMSSM model with non-minimal soft terms in the Higgs sector. 
We view this as an effective theory whose UV cutoff is the lowest decoupling scale among the  $M, M_D, M_T, M_F, M_P, M_E$ and
$M_S$. For the rest of this paper we assume that this is $M$, i.e. the RG running always commences at this scale. After choosing the input, 
we run all parameters down to the EW scale using one-loop RGEs. The relevant threshold effect comes from the heavy states in the Higgs sector 
with  mass $m_A$. At this scale the heavy $SU(2)$ Higgs doublet $H_d$ is integrated out. The bino, the Higgsinos and the left handed sleptons all live at the 
weak scale $m_Z$. For the rest of the sparticle spectrum we assume a common decoupling mass scale $m_{SUSY}$. After minimizing the scalar Higgs potential once, we iterate 
the procedure, each time adjusting the values of the input parameters in order to correctly reproduce the top mass as well as the mass of the $Z$-boson. To be 
exact we impose the following two constraints at the weak scale:

\smallskip
1. We need to have $\,v = \sqrt{v_d^2 + v_u^2} = 174\,$ GeV in order to reproduce the correct value of the $Z$-boson mass $\,m_Z\,$ (where as usual 
$\,v_u = \langle H_u^0\rangle\,$ and $\,v_d = \langle H_d^0 \rangle\,$ denote the vacuum expectation values of the neutral 
components of the two Higgs doublets).

\smallskip
2. The low-energy value of the top Yukawa coupling $\,y_t\,$ should correctly reproduce the top quark mass, i.e. $\,m_t\,=\,y_t(M_{\rm EW}) v_u$
\footnotemark[1] \footnotetext[1]{Here $m_t$ stands for the running top mass which differs from the pole mass $m_t^{pole}$. To order $\alpha_s$ the 
relation between the two is given by $m_t^{pole} = m_t(m_t) \left( 1 + \frac{4\alpha_s}{3\pi} \right)$.}

\section{Origin and composition of the effective $\mu$ term}\label{mu}

It has been known for quite some time that the minimal GMSB extension of the NMSSM does not lead to phenomenologically viable spectra 
due to the small value of the induced $\mu$ term. As noted already in \cite{deGouvea:1997cx} this problem can be restated in terms of 
$\det \mathcal{M}^2_{\rm CP-even}$, the determinant of the CP-even squared mass matrix in the Higgs sector. Specifically one 
can show that large values of the effective SUSY breaking scale, which are necessary in order to satisfy the current LHC bound 
on the gluino, can lead to a negative $\,\,\det \mathcal{M}^2_{\rm CP-even} < 0$. 

The origin of this result can be understood analytically if one  examines in more detail the following  minimization condition in the Higgs sector:
\bea
2\,\frac{\kappa^2}{\lambda^2} \, ( \,\lambda^2 s^2 \,) \, = \,\lambda v^2 (\kappa \sin 2\beta - \lambda) - m_N^2 + A_{\lambda} \lambda \, v^2 \, \frac{\sin 2\beta}{2 s} + 
\kappa A_{\kappa} \, s \,\,. \label{minimization}
\eea 
Here $\,s = \langle N \rangle\,$ is the VEV of the scalar component of $N$. 
One can argue that in minimal gauge mediation all terms on the right hand side remain relatively small. This applies in particular to the 
$m_N^2$ soft mass which is zero at the messenger scale and remains small all the way down to $\,m_{\rm EW}\,$ due to its small $\beta$-function.
As we will see this is the key ingredient that changes within the lopsided setting. For now let us see what the implications of the aforementioned 
observation are. We will argue that eq.(\ref{minimization}) imposes an upper bound on the $\mu$ term in MGM models (it restricts this mass term to several GeV). For the sake of contradiction
assume that $\,\mu_{\rm eff, \,MGM} = \lambda s\,$ can be of the order $\mathcal{O}(100)$ GeV. In this case the left hand side of (\ref{minimization}) will be unacceptably large
unless one imposes $\lambda \gg \kappa$. In order to see why this relation is problematic, let us use the following approximation for $\det \mathcal{M}_{\rm CP-even}^2$ 
which retains only the terms with the highest power of $\mu$ (see \cite{deGouvea:1997cx}):
\bea
\det \mathcal{M}_{\rm CP-even}^2 \simeq {\rm const.} \, \mu^4 \, \left(-4 k \lambda^4 + 2 \kappa^3 g^2 + 2 \kappa^3 g^2 \,\cos (4\beta) + 8\kappa^2 \lambda^3 \,\sin(2 \beta)
\label{matapprox} 
\,\right) 
\eea
Given that $\lambda \gg \kappa$, it is clear that the first term dominates, leading to a negative $\det \mathcal{M}_{\rm CP-even}^2 < 0$. To make things worse, the small 
value of the $\kappa$ coupling implies a nearly massless ``Goldstone'' mode in the low-energy spectrum (recall that in the MGM version of the NMSSM $\,\frac{\kappa}{3}\,N^3\,$ 
is the only coupling which breaks Peccei-Quinn explicitly). 

In the lopsided NMSSM the aforementioned problem is no longer present because one generically expects that the values of $\kappa$ and $\lambda$ will be of the same order of magnitude. 
Accordingly, contributions from the last three \textit{positive} terms in eq.(\ref{matapprox}) become comparable to and can cancel out the negative contribution from the first term. 
In this case the determinant of the Higgs mass matrix becomes positive. 

Let us now try to understand the theoretical underpinning behind the statement that the $\lambda$ and $\kappa$ couplings are expected to be of the same order of magnitude. To this end 
we shall once again take a look at the minimization condition (\ref{minimization}). Previously we argued that all terms on the right hand side are very small compared to $\lambda s$. 
This applies in particular to the singlet soft mass $m_N^2$ which is practically zero at the messenger scale and remains very small all the way down to the EW scale due to the small 
$\beta$-function. In the lopsided NMSSM this picture changes due to the non-typical RG running of $m_N^2$, an effect which is similar to the one we encountered in the slepton and the 
squark sectors. To make this last statement more precise, lets take a look at the one-loop RGE for the singlet soft mass $m_N^2$:
\bea
16\pi^2 \, \frac{d}{dt} \,m_{N}^2 = 4 \lambda^2 \left(\,m_{H_u}^2 + m_{H_d}^2 + m_N^2 + 4 A_{\lambda}^2 \,\right) \, + \,4 \kappa^2 \left( 3 m_N^2 + 4 A_{\kappa}^2 \right)
\eea
Note that the dominant $\,m_{H_d}^2\,$ term generates a large negative $m_{N}^2$ along the RG trajectory. In particular eq.(\ref{minimization}) can be satisfied without invoking 
$\lambda \gg \kappa$ which solves the aforementioned determinant problem. Additionally, the large $\,m_{N}^2\,$ triggers a VEV for the singlet $N$ which translates into 
a sizeable contribution to the $\mu$ term. Last but not least, the moderately large value of $\kappa$ implies that PQ is no longer an approximate symmetry of the Lagrangian. 
Hence the quasi-Goldstone mode in the Higgs sector is avoided in a completely natural manner.

\section{Low-energy spectrum}

\subsection{Higgs sector}\label{higgs}

\begin{figure}
\caption{The radiative corrections to the Higgs mass as a function of $A_t$ and $m_{SUSY}$.\label{fig1}}
\begin{center}
\includegraphics[width=10cm]{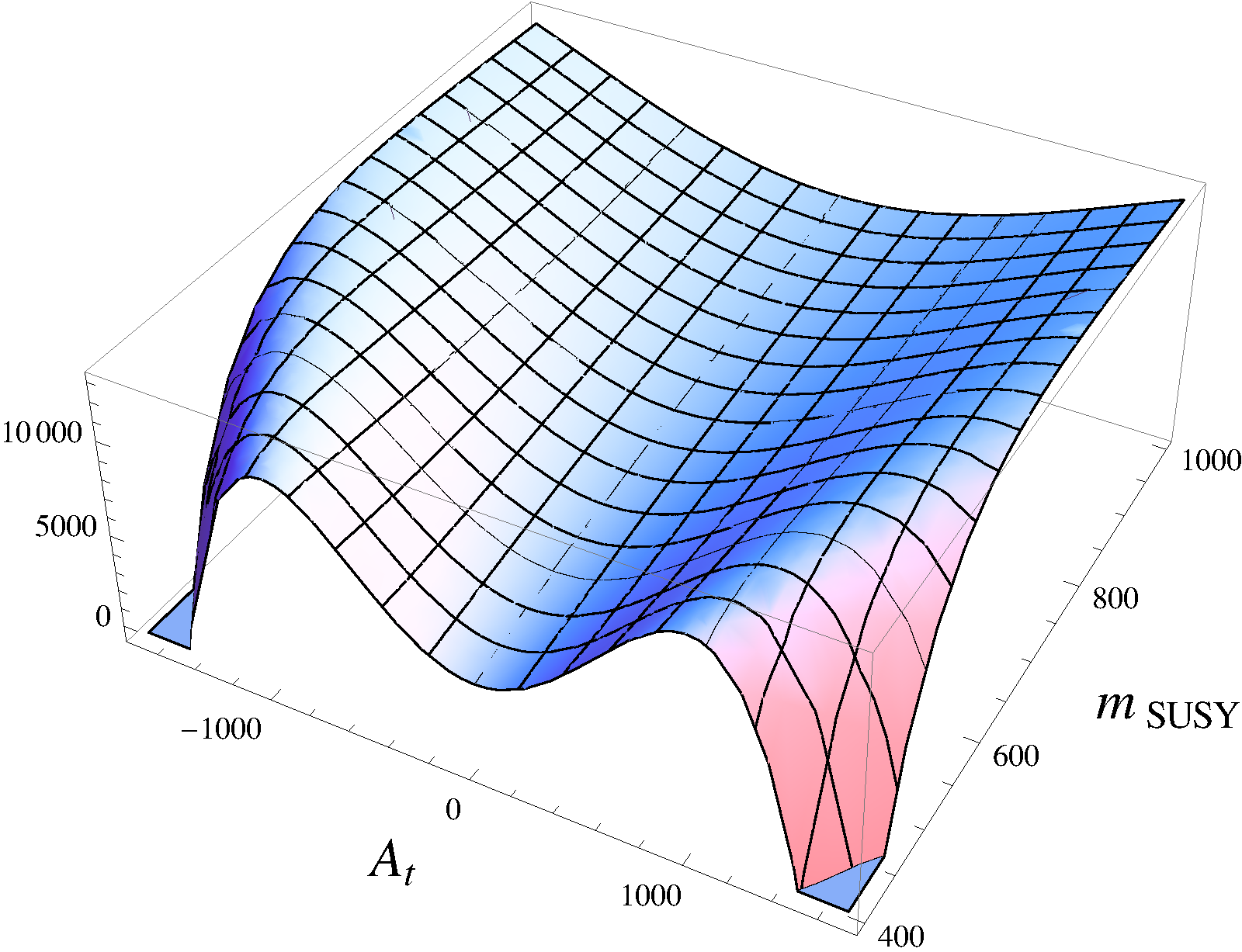}
\end{center}
\end{figure}

The crucial issue in this section is the mass of the lightest CP-even neutral Higgs boson. Our discussion is based 
on the approximate one-loop formula \cite{Ellwanger:2009dp}
\bea\label{higgsmass}
m_{h^0}^2  \, &=& \, m_Z^2 \cos^2 (2 \beta) \, + \, \lambda^2 v^2 \sin^2(2 \beta) \, \, - \,\frac{\lambda^2}{\kappa^2} \, v^2 \, (\,\lambda - \kappa \sin (2\beta)\,)^2 + \nonumber\\
&+&\frac{3 m_{t}^4}{4\pi^2 v^2} \, \left(\,\ln\left(\, \frac{m_{SUSY}^2}{m^2_t} \,\right) + \frac{X_t^2}{m_{SUSY}^2}\, \left(1- \frac{X_t^2}{12 \,m_{SUSY}^2} \right)  \,\right)\,\,.
\eea
where $X_t = A_t - \mu \cot \beta\,$ is the stop mixing parameter and  $\,m_{SUSY} \equiv \sqrt{m_{\tilde{1}} \, m_{\tilde{2}}}\,$ stands for the SUSY scale. Here we use eq.(\ref{higgsmass}) only for 
illustrative purposes, the actual calculation is done using NMSSMTools \cite{Ellwanger:2004xm, Ellwanger:2006rn}. Let us first analyze the radiative 
correction specified by the second line in (\ref{higgsmass}). To this end we plot the magnitude of this correction as a function of $\,A_t$ and $m_{SUSY}\,$ in Figure \ref{fig1}. The symmetric ridges on the 
left and right hand sides denote the phenomenologically preferred region where the one-loop radiative correction is maximized.

Very recently the authors of \cite{Draper:2011aa} have performed a similar analysis focussing on the MSSM piece of eq.(\ref{higgsmass}). It was argued that in order to lift the Higgs mass up to $125$ GeV one 
needs a large soft trilinear term $A_t > 1$ TeV, unless of course one is prepared to tolerate superheavy stops ($\sim 5-10$ TeV) in the theory. This is in agreement with our plot which shows that the radiative correction is 
maximized for large values of the soft trilinear term. For the remainder of the paper we focus on those regions of parameter space with $\,|A_t| > 1\,$ TeV and identify $X_t \approx A_t$.

Next we  discuss the NMSSM specific contributions to the tree-level $m_{h^0}^2$. Since $\tan \beta \gg 1\,$ we infer that  $\,\lambda^2 v^2 \sin^2(2 \beta) \approx 0$. The negative tree-level contribution
due to the mixing with the singlet requires more detailed analysis. In the following we will look at several points in parameter space with $\,|A_t| > 1\,$ TeV and we will evaluate its magnitude. In order 
to fix the notation we introduce the three rotation angles $\,\Theta_T$, $\Theta_P$, $\Theta_S\,$ as well as the proportionality factors $\,\xi_T$, $\xi_P$, $\xi_S\,$ corresponding to the last 
three lines in eq.(\ref{fullUV}):
\bea
\xi_T = 1.0 \quad\quad\,\,\,   \xi_S = 1.0 \quad\quad \tan \Theta_P = 1.0 
\eea
We refrain from specifying $\tan \Theta_T$ and $\tan \Theta_S$ since their values will not affect any of the mass terms in the Higgs sector. Next we fix the following mass scales:
\bea
&&  M_T/M_D = 1.1  \quad\quad\quad M_P/M_F = 1.1  \quad\quad\quad M_S/M_E = 1.1 \\ 
&&  M_D = 500 \,\,\,\mbox{TeV} \quad\quad\,\, M_F = 500 \,\,\, \mbox{TeV} \quad\quad\,\,M_E = 500 \,\,\, \mbox{TeV}  \\
&&  \quad\quad\quad\quad\quad \,\,\,\,\, \quad\quad\quad\,\,\Lambda_P = 24 \,\,\,\mbox{TeV} \quad\quad\quad\,\Lambda_S = 80 \,\,\,\mbox{TeV} \\
&&  \quad\quad\quad\quad\quad\quad \, \quad\quad\quad\Lambda_F = 24 \,\,\,\mbox{TeV} \quad\quad\quad\,\Lambda_E = 80 \,\,\,\mbox{TeV} 
\eea

The three models (or more precisely the three points in parameter space) we consider are specified by varying the residual input parameters (Table \ref{tab1}).
\begin{table}[ht]
\begin{center}
\caption{Input parameters at the messenger scale $M$ \label{tab1}}
\begin{tabular}{|l|c|c|c|}
\hline
\quad\,              & Point 1     & Point 2    & Point 3  \\
\hline
$M$              &  $10^5$ GeV    &   $10^5$ GeV  &   $10^5$ GeV  \\
\hline
$\xi_{P}$        &   6.75       &    6.75     &   6.75    \\
\hline
$\Lambda$        &   50.0 TeV    &   50.0  TeV      &   50.0  TeV    \\
\hline
$\Lambda_T$      &   63.7 TeV    &   63.7  TeV      &   63.7  TeV  \\
\hline
$\Lambda_D$      &   10.4  TeV  &   10.4 TeV     &   10.4  TeV  \\
\hline
$\kappa$         &   0.18        &   0.16     &   0.19    \\
\hline
$\lambda$        &   0.21         &  0.20     &   0.20    \\
\hline
$\lambda_{u,T}$   &   0.53      &   0.55     &   0.48    \\
\hline
$\lambda_{d,T}$   &   2.29      &   2.24     &   2.51    \\
\hline
$\lambda_{u,P}$  &   0.28      &   1.24     &   1.03    \\
\hline
$\lambda_{d,P}$  &   0.28      &   1.24     &   1.03    \\
\hline
$\lambda_{u,S}$  &   2.00      &   2.00     &   2.00    \\
\hline
$\lambda_{d,S}$  &   0.00      &   0.00     &   0.00    \\
\hline
\end{tabular}
\end{center}
\end{table}

The values of all Yukawa couplings are given at the messenger scale $M$. In Table \ref{tab2} we evaluate the mixing effect for the three 
points given in Table \ref{tab1}.

\begin{table}[ht]
\begin{center}
\caption{Mixing effect due to the presence of a gauge singlet $N$\label{tab2}}
\begin{tabular}{|l|c|c|c|}
\hline
\quad\,                                                                            & Point 1     & Point 2    & Point 3  \\
\hline
$-\lambda^2/\kappa^2 \, \times \, v^2 \, (\,\lambda - \kappa \sin (2\beta)\,)^2$   &  -6892  \,GeV$^2$    &   -7593  \,GeV$^2$   &   -5269  \,GeV$^2$  \\
\hline
$m_Z^2 \cos^2 (2 \beta) \, + \, \lambda^2 v^2 \sin^2(2 \beta)$                     &   8190  \,GeV$^2$     &   8188  \,GeV$^2$    &   8188  \,GeV$^2$   \\
\hline
\end{tabular}
\end{center}
\end{table}

Clearly, the mixing effect is unacceptably large for every single one of the points in parameter space we have investigated. 
The problem can be traced back to the fact that  $\,\kappa < \lambda$ and $\tan \beta \gg 1$. From these we can deduce that 
\bea
\frac{\lambda}{\kappa} \, v \, (\,\lambda - \kappa \sin (2\beta)\,) \approx \frac{\lambda^2}{\kappa} \, v \,\, \quad\quad (\lambda \gg \kappa \sin 2\beta)
\eea
i.e. the negative tree-level contribution is significant. In the following we will discuss a mechanism which will allow us to largely suppress the $\,\,\frac{\lambda^2}{\kappa^2} \, v^2 \, (\,\lambda - \kappa \sin (2\beta)\,)^2\,\,$ term. 
The argument is based on the three minimization conditions in the Higgs sector:
\bea
&\mu_{\rm eff}^2 \equiv (\mu + \lambda s)^2 = -\,\frac{M_Z^2}{2} \, + \, \frac{m_{H_d}^2-m_{H_u}^2 \tan^2 \beta}{\tan^2 \beta - 1} \label{min1}\\ 
\noalign{\medskip}
&B_{\mu,\rm eff} \equiv  (\mu + \lambda s)(A_{\lambda} + \kappa s) + B_{\mu} = \left(\,m_{H_u}^2 + m_{H_d}^2 + 2\lambda^2 s^2 + 2\lambda^2 v^2 \,\right) \,\frac{\sin 2\beta}{2} \label{min2} \\
\noalign{\medskip}
&2\,\frac{\kappa^2}{\lambda^2} \, ( \,\lambda^2 s^2 \,) \, = \,\lambda v^2 (\,\kappa \sin 2\beta - \lambda) - m_N^2 + A_{\lambda} \lambda \, v^2 \, \frac{\sin 2\beta}{2 s} + 
\kappa A_{\kappa} \, s  \label{min3}
\eea
Assume that we start in a region of parameter space where $\lambda > \kappa$. This would be the case if, for example, we picked one of the points listed in Table \ref{tab3}.
In the following we will specify a procedure which will allow us to push our model into a region of parameter space where $\kappa \gg \lambda$. In this case the two terms $\lambda$
and $\,\kappa \sin(2\beta)\,$ will be comparable in size and will cancel each other out leading to a highly suppressed or altogether vanishing mixing effect.

The idea of our construction is very simple -- we hold the effective $\mu_{\rm eff} = \mu + \lambda s$ term constant, while simultaneously increasing the tree-level $\mu$ piece (note that 
$\mu_{\rm eff}$ is comprised of two pieces, with $\lambda s$ being the dynamically generated one). This can be done for instance by
increasing the $\,\xi_{P}\,$ parameter. Note that this procedure does not affect the $m_{H_u}^2$ and $m_{H_d}^2$ soft masses
meaning that the first minimization condition (\ref{min1}) remains identically satisfied at every step of our construction. The third minimization condition given by eq.(\ref{min3}) is 
somewhat more problematic. Note that by increasing $\mu$ we automatically decrease the $\lambda s$ contribution to $\mu_{\rm eff}$. In particular the left hand side of eq.(\ref{min3}) 
becomes smaller in size while the right hand side remains more or less unaffected. This is easily deduced by observing that the main contribution to the r.h.s. comes from the 
$m_{N}^2$ soft mass term whose value is insensitive to variations of $\mu$ or $\lambda s$. Thus in order to preserve the validity of eq.(\ref{min3}) we need to increase the $\kappa$ coupling.
We can then iterate this procedure until we get $\kappa \gg \lambda$. 

The only open issue that remains is the validity of eq.(\ref{min2}). Clearly the procedure we just described affects both the left and the right hand side of this equation and there is 
no a priori reason why this identity should remain valid after the model parameters have been modified. Here we make use of the fact that $B_{\mu}$ transforms 
under $U(1)_R$ phase rotations while the other mass terms in the Lagrangian do not (see also \cite{DeSimone:2011va}). In particular we can use the $U(1)_R$ symmetry to forbid a $B_{\mu}$ term 
altogether. To put it differently -- the overall size of $B_{\mu}$ must be a free parameter which should depend on the value of some appropriately chosen coupling in the high-energy model (cf. 
the discussion in section 3). Hence we can always assume that $B_{\mu}$ is chosen in such a manner that eq.(\ref{min2}) is identically satisfied.

In Table \ref{tab3}, we list several points in parameter space with $\,\kappa \gg \lambda\,$ or $\,\kappa > \lambda$.

\begin{table}[ht]
\begin{center}
\caption{Five points in parameter space with a large $\kappa$ coupling\label{tab3}}
\begin{tabular}{|l|c|c|c|c|c|}
\hline
\quad\,           & Point 4     & Point 5       & Point 6      & Point 7      & Point 8   \\
\hline
$M$              &  $10^5$ GeV  &   $10^5$ GeV  &   $10^5$ GeV &   $10^5$ GeV  &   $10^5$ GeV \\
\hline
$\xi_{P}$         &   6.75     &    6.75       &   5.45       &    6.75       &   5.45     \\
\hline
$\Lambda$        &   60.0 TeV  &   50.0  TeV    &  50.0 TeV    &   50.0  TeV  &   50.0  TeV  \\
\hline
$\Lambda_T$      &   67.9 TeV  &   63.7  TeV    &   57.6  TeV  &   67.2  TeV &   58.8 TeV \\
\hline
$\Lambda_D$      &   11.0 TeV  &   10.4 GeV    &   9.4  GeV   &   10.9 GeV   &   9.55  GeV   \\
\hline
$\kappa$         &   0.58      &   0.58         &   0.56      &   0.58        &   0.56  \\
\hline
$\lambda$        &   0.32      &   0.32          &   0.12     &   0.32       &   0.12   \\
\hline
$\lambda_{u,T}$  &   0.66      &   0.58         &   0.65       &   0.57        &   0.59     \\
\hline
$\lambda_{d,T}$  &   2.91      &   2.22         &    2.38      &   2.18        &    2.83   \\
\hline
$\lambda_{u,P}$  &   1.20      &   1.30         &   1.45       &   1.27        &   2.05   \\
\hline
$\lambda_{d,P}$  &   1.28      &   1.30         &   1.45       &   1.27        &   2.05    \\
\hline
$\lambda_{u,S}$  &   2.15      &   2.00         &   2.00       &   1.97        &   1.75   \\
\hline
$\lambda_{d,S}$  &   0.00      &   0.00         &   0.00       &   0.00        &   0.00     \\
\hline
\end{tabular}
\end{center}
\end{table}

For all points in Table \ref{tab3} we have chosen a large $\tan \beta = 15$. The mass of the lightest Higgs boson as well as the value and composition 
of the effective $\mu_{\rm eff}$ term are given in Table \ref{tab4}. Point 4 is an example of a model in which the effective $\mu_{\rm eff}$ term arises predominantly 
through the NMSSM mechanism. Conversely, for points 6 and 8 the tree-level contribution $\mu$ dominates. In fact the models represented by 6 and 8 can 
be viewed as the MSSM limit of the lopsided NMSSM since the effective mass term for the Higgsinos is almost entirely composed of the tree level 
$\mu$ term generated by the Higgs-messenger couplings.

\begin{table}[ht]
\begin{center}
\caption{Higgs mass and composition of the effective $\mu$ term \label{tab4}}
\begin{tabular}{|l|c|c|c|c|c|}
\hline
                  &  Point 4               & Point 5     & Point 6       & Point 7        & Point 8     \\
\hline
 $m_{h^0}$        &   $124$ GeV        &  $124$ GeV &   $124$ GeV  &   $124$ GeV   &    $124$ GeV             \\
\hline
 $\mu_{\rm eff}$  &   $314$ GeV        & $262$ GeV  &  $140$ GeV  &  $268$ GeV   &  $262$ GeV  \\
\hline
 $\,\,\,\mu$      &   $100$ GeV        & $111$ GeV  &  $118$ GeV  &  $105$ GeV   &  $236$ GeV   \\
\hline
 $\,\,\lambda s$  &   $214$ GeV        & $151$ GeV   &  $22$ GeV   &  $163$ GeV    &  $26$ GeV   \\
\hline
\end{tabular}
\end{center}
\end{table}

\subsection{Slepton and squark sector}\label{slepton}

The slepton sector of the lopsided NMSSM exhibits some peculiar features which can be traced back to the large value of the soft $m_{H_d}^2$ mass parameter. 
To be more precise, $m_{H_d}^2$ can have a significant impact on the running of the right and left handed soft slepton masses due to the presence of the
$\,\frac{6}{5}\,Y_{\tilde{f}}\, g_1^2 S\,$ term in the $\beta$-function (see Appendix A). The aforementioned contribution is related to the induced hypercharge Fayet-Illiopoulus term in 
the effective action. Note that in the MSSM the impact of this term on the RG running is comparatively small. However, its significance increases dramatically once the $m_{H_d}^2$ is allowed 
to attain very large values. It is also important to note that $\,\frac{6}{5}\,Y_{\tilde{f}}\, g_1^2 S\,$  contributes with opposite signs to the $\beta$-functions 
of the right handed and left handed sleptons. Specifically, it produces a negative contribution to soft masses $m_{\tilde{L}_L}^2$ of the left-handed sleptons along the RG trajectory and a positive one 
to the masses $m_{\tilde{L}_R}^2$ of their right handed counterparts. As we will see shortly, this effect often leads to an atypical hierarchy $m_{\tilde{L}_R}^2 > m_{\tilde{L}_L}^2$ at the electroweak 
scale and can therefore reverse the ordering of the right and left handed slepton masses.

\begin{table}[ht]
\begin{center}
\caption{Low-energy spectrum in the slepton sector\label{tab5}}
\begin{tabular}{|l|c|c|c|c|c|}
\hline
                           &  Point 4       & Point 5   & Point 6     & Point 7   & Point 8       \\
\hline
  $m_{\tilde{e}_R}$       &  $499$ GeV    & $361$ GeV &  $371$ GeV  & $383$ GeV &  $414$ GeV   \\
\hline
  $m_{\tilde{e}_L}$       &  $327$ GeV      & $307$ GeV &  $305$ GeV  & $297$ GeV &  $273$ GeV  \\
\hline
 $m_{\tilde{\nu}_e}$      &  $318$ GeV     & $297$ GeV &  $295$ GeV  & $287$ GeV &  $261$ GeV  \\
\hline
   $m_{\tilde{\mu}_R}$    &  $499$ GeV     & $361$ GeV &  $371$ GeV  & $383$ GeV &  $414$ GeV    \\
\hline
  $m_{\tilde{\mu}_L}$     &  $327$ GeV     & $307$ GeV &  $305$ GeV  & $297$ GeV &  $273$ GeV  \\
\hline
 $m_{\tilde{\nu}_{\mu}}$  &  $318$ GeV     & $297$ GeV &  $295$ GeV  & $287$ GeV &  $261$ GeV  \\
\hline
 $m_{\tilde{\tau}_1}$     &  $260$ GeV     & $270$ GeV &  $263$ GeV  & $256$ GeV &  $205$ GeV    \\
\hline
  $m_{\tilde{\tau}_2}$    &  $419$ GeV     & $315$ GeV &  $308$ GeV  & $328$ GeV &  $334$ GeV  \\
\hline
 $m_{\tilde{\nu}_{\tau}}$ &  $251$ GeV      & $267$ GeV &  $254$ GeV  & $249$ GeV  &  $194$ GeV   \\
\hline
\end{tabular}
\end{center}
\end{table}

\begin{table}[ht]\label{squarks}
\begin{center}
\caption{Low-energy spectrum in the squark sector\label{tab6}}
\begin{tabular}{|l|c|c|c|c|c|}
\hline
                          &  Point 4       & Point 5   & Point 6   & Point 7   & Point 8        \\
\hline
 $m_{\tilde{u}_R}$        &  $1552$ GeV     & $1278$ GeV &  $1278$ GeV   & $1303$ GeV &  $1264$ GeV  \\
\hline
  $m_{\tilde{u}_L}$       &  $1661$ GeV     & $1356$ GeV &  $1254$ GeV   & $1385$ GeV &  $1356$ GeV \\
\hline
  $m_{\tilde{d}_R}$       &  $1613$ GeV     & $1313$ GeV &  $1310$ GeV   & $1343$ GeV &  $1316$ GeV \\
\hline
  $m_{\tilde{d}_L}$       &  $1663$ GeV     & $1358$ GeV &  $1356$ GeV   & $1387$ GeV &  $1358$ GeV   \\
\hline
   $m_{\tilde{s}_R}$      &  $1613$ GeV     & $1313$ GeV &  $1310$ GeV   & $1343$ GeV &  $1316$ GeV \\
\hline
   $m_{\tilde{s}_L}$      &  $1663$ GeV     & $1358$ GeV &  $1356$ GeV   & $1387$ GeV &  $1358$ GeV \\
\hline
  $m_{\tilde{c}_R}$       &  $1552$ GeV     & $1278$ GeV &  $1278$ GeV   & $1303$ GeV &  $1264$ GeV    \\
\hline
   $m_{\tilde{c}_L}$      &  $1661$ GeV     & $1356$ GeV &  $1354$ GeV   & $1387$ GeV &  $1356$ GeV   \\
\hline
 $m_{\tilde{t}_{1}}$      &  $1033$ GeV     & $804$ GeV &  $777$ GeV   & $819$ GeV    &  $814$ GeV  \\
\hline
 $m_{\tilde{t}_2}$        &  $1491$ GeV     & $1241$ GeV &  $1232$ GeV   & $1254$ GeV &  $1239$ GeV    \\
\hline
  $m_{\tilde{b}_1}$       &  $1435$ GeV     & $1168$ GeV &  $1158$ GeV   & $1188$ GeV &  $1169$ GeV  \\
\hline
 $m_{\tilde{b}_{2}}$      &  $1541$ GeV     & $1266$ GeV &  $1263$ GeV   & $1292$ GeV &  $1254$ GeV \\
\hline
\end{tabular}
\end{center}
\end{table}

In Table \ref{tab5} we list the low-energy spectrum for the same points in parameter space that we discussed at the end of section 3. The effect which reverses the 
ordering of the left-handed and right-handed sleptons is clearly visible both in models with low and intermediate scale gauge mediation. 

In the following we will argue that the $m_{H_d}^2$ soft mass and the corresponding $\lambda_d$ coupling cannot attain arbitrarily large values. One 
constraint arises from the requirement that the slepton masses should remain positive. Since the slepton particles are not charged under the $SU(3)$, their 
soft masses at the messenger scale are relatively small. Therefore the negative contribution arising from the $\,\frac{6}{5}\,Y_{\tilde{f}}\, g_1^2 S\,$ 
Fayet-Illiopoulus term can, at least in principle, make some of the slepton masses negative. Clearly, the tau sneutrinos are the lightest particles in the 
slepton sector. Imposing the condition $m_{L_3}^2 > 0$ at the weak scale is sufficient to guarantee positivity of $m_{\tilde{\nu}_{\tau}}$ (and, a 
fortiori, of all slepton masses). For the models under consideration this does not introduce an upper bound on $\lambda_d$ -- even for non-perturbative
values of $\,\lambda_d\,$ at the messenger scale, the $\,m_{L_3}^2\,$ soft mass term remains positive at the weak scale. An actual bound on  $\,\lambda_d\,$
can be obtained by requiring that this coupling remains perturbative up to the GUT scale. From the one-loop RG equation for $\lambda_d$
\bea
\frac{d\lambda_d^2}{dt} \, = \,\frac{\lambda_d^2}{8\pi^2} \,\left(\,4\lambda_d^2 - \frac{3}{5}g_1^2 - 3g_2^2 \,\right)\,,\quad t = \log Q
\eea                      
one can deduce e.g. $\lambda_d \lesssim 1.0$ for the models with $M = 10^5$ GeV and $\lambda_d \lesssim 1.1$ for the models with $M = 10^8$ GeV. This is 
clearly incompatible with the values we have picked. As pointed out in \cite{Csaki:2008sr} larger values for the $\lambda_d$ coupling can be obtained by
assuming that the SUSY breaking sector is strongly coupled.

There are two important effects which characterize the squark sector of the lopsided NMSSM. The first is related to the hypercharge Fayet-Illiopoulus contribution to the 
$\beta$-functions of the left and right handed squarks. Note that the FI term contributes with opposite signs to the $\beta$-functions of $\,m_U^2\,$ and 
$\,m_D^2$. Since squarks are charged under the $SU(3)$ group, the RGEs are dominated by the gluino contribution $\,\propto g_3^2 \,|M_3|^2\,$ and the FI 
effect is not that prominent so as to reverse the ordering of the up and down squark masses. What it does, however, is to decrease the mass gap between right and left 
handed sparticles. The other effect is related to the large value of the $A_t$ soft trilinear term. Looking at the tree-level stop mass matrix   
\bea
\small
\mathbf{m}_{\rm \tilde{t}}^2  =        \left( \begin{array} {cc}
                    m_{Q_3}^2 + m_t^2 - \left(\,\frac{1}{2} + \frac{2}{3}\,\sin^2\theta_W \,\right)\,m_Z^2 \cos 2\beta    &  m_t \left( A_t - \mu \cot \beta \right)      \\
                    m_t \left( A_t - \mu \cot \beta \right)             & m_{U_3}^2  +  m_t^2 + \frac{2}{3}\,\sin^2\theta_W \,m_Z^2 \cos 2\beta   
                    \end{array} \right) \label{matrixstop}
\eea
it is immediately clear that the off-diagonal terms are large, leading to a significant splitting between the two mass eigenstates $\,m_{\tilde{t}_1}\,$ and $\,m_{\tilde{t}_2}$.
For the sbottom mass eigenstates $\,m_{\tilde{b}_1}\,$ and $\,m_{\tilde{b}_2}\,$ the splitting is negligible. This is most easily seen by considering the respective mass matrix 
\bea
\small
\mathbf{m}_{\rm \tilde{b}}^2  =        \left( \begin{array} {cc}
                    m_{Q_3}^2 + m_b^2 - \left(\,\frac{1}{2} - \frac{1}{3}\,\sin^2\theta_W \,\right)\,m_Z^2 \cos 2\beta    &  m_b \left( A_b - \mu \cot \beta \right)      \\
                    m_b \left( A_b - \mu \cot \beta \right)             & m_{D_3}^2  +  m_b^2  -  \frac{1}{3}\,\sin^2\theta_W \,m_Z^2 \cos 2\beta   
                    \end{array} \right) \label{matrixstop2}
\eea
and noting that $\,m_b A_b \ll m_t A_t$. The full spectra are listed in Table \ref{tab6}.

\subsection{Charginos, neutralinos, gluino}\label{higgsinos}

The gauge singlet $N$ adds one extra degree of freedom in the neutralino sector. To make this more precise let us 
introduce the fermionic component $\psi_N$ of the superfield $N$. Since $N$ is a gauge singlet, the $\psi_N$
Weyl spinor is itself uncharged under the SM gauge group and we will refer to it as a singlino. 
The $5 \times 5$ neutralino mass matrix reads in the basis $(\tilde{B}^0, \tilde{W}_3^0, \tilde{H}_d^0, \tilde{H}_u^0, \psi_N)\,$ 
(cf. \cite{Ellwanger:2009dp}):
\bea
\small
\mathcal{M}_{\tilde{\chi}^0}  =        \left( \begin{array} {ccccc}
                M_1     & 0   &  -m_Z \sin\theta_W \cos \beta  & m_Z \sin\theta_W \sin \beta    &  0    \\
                0       & M_2 &   m_Z \cos\theta_W \cos \beta  & -m_Z \cos\theta_W \sin \beta   &  0    \\   
      -m_Z \sin\theta_W \cos \beta    & m_Z \cos\theta_W \cos \beta   & 0   & -\mu_{\rm eff}   &  - \lambda v_u   \\ 
       m_Z \sin\theta_W \sin \beta    & -m_Z \cos\theta_W \sin \beta  & -\mu_{\rm eff} &  0       &  - \lambda v_d       \\
           0          & 0  &  - \lambda v_u  & - \lambda v_d   &  2 \kappa \,s     \\
                    \end{array} \right)\nonumber \label{neutralino}
\eea
The $4 \times 4$ mass matrix in the chargino sector is identical to its MSSM counterpart. Overall we get the spectra for the 
five points in parameter space shown in Table \ref{tab7}.

\begin{table}[ht]
\begin{center}
\caption{Charginos and neutralinos\label{tab7}}
\begin{tabular}{|l|c|c|c|c|c|}
\hline
\mbox{Sparticle}            &  Point 4  &  Point 5  &  Point 6 &  Point 7 &  Point 8 \\
\hline
 $\,\quad m_{\chi^0_1}$     &    167    &    103    &    157   &    166  &    112  \\
\hline
 $\,\quad m_{\chi^0_2}$     &    228    &   - 141\,\,    &   178    &    206   & - 146  \\
\hline
 $\,\quad m_{\chi^0_3}$     &  - 229 \,\,&   143   &   - 224\,\, &   232    &    210  \\
\hline
 $\,\quad m_{\chi^0_4}$     &    399    &    210    &    226  &   - 232\,\,   &    223   \\
\hline
 $\,\quad m_{\chi^0_5}$     &    462    &    394     &    390  &    391   &    394  \\
\hline
 $\,\quad m_{\chi^{\pm}_1}$ &   205     &    124    &    200   &    208  &    129  \\ 
 \hline
 $\,\quad m_{\chi^{\pm}_2}$ &   399     &    394     &    389   &    390   &    394  \\  
\hline
\end{tabular}
\end{center}
\end{table}

The neutralino $\chi^0_1$ which arises from the mixing with the singlet is the lightest particle in the chargino/neutralino sector.
A quick glance at the values in Table \ref{tab7} reveals that the size of $m_{\chi^0_1}$  depends on the value of the effective $\mu$ term:
In models with larger $\mu_{\rm eff}$ the respective mass $m_{\chi^0_1}$ is larger and conversely -- for smaller values of $\mu_{\rm eff}$
within the range of $\,\sim 100\,$ GeV the respective neutralino mass is small leading to models with $\chi^0_1$ as NLSP. 

However, the lightest neutralino is not always NLSP. Depending on the input parameters the stau sneutrino may become lighter. An important effect, which determines the mass 
 $m_{\tilde{\nu}_{\tau}}\,$, is the size of the induced FI term in the effective action, or in other words the size of the $\,m_{H_d}^2\,$ soft mass. The $\,m_{H_d}^2\,$
parameter affects the size of $m_{\tilde{\nu}_{\tau}}$ indirectly through the $\beta$-funcion of $m_{L_3}^2$, by decreasing the value of $m_{L_3}^2$ along the RG trajectory.

Overall we deduce that models with a large effective $\mu_{\rm eff}$ term favour $m_{\tilde{\nu}_{\tau}}$ as the NLSP whereas in models with a small $\mu_{\rm eff}$ there is a
tension between the $m_{\tilde{\nu}_{\tau}}$ and $m_{\chi^0_1}$ masses. Point 8 is an example of a model with stau sneutrino NLSP, whereas Point 6 has $\chi^0_1$ as NLSP.

\section{Collider phenomenology}

In theories with gauge mediated supersymmetry breaking the lightest supersymmetric particle (LSP) is the gravitino. The lopsided NMSSM
makes no exception to this rule. In the lopsided NMSSM the next to lightest supersymmetric particle (NLSP) can be either the lightest neutralino 
$\chi_1^0$ or the stau sneutrino $\tilde{\nu}$, depending on how large the effective $\mu$ term is. The crucial difference between the lopsided 
NMSSM and its MSSM counterpart is the fact that in the former case the effective $\mu$ term is not as strictly bounded from above.

After its production any supersymmetric state decays until the NLSP state is reached. The decay length of the NLSP is given by 
\bea
L \approx 10^{-2} \,{\rm cm} \left(\frac{100 \, {\rm GeV}}{m_{\rm NLSP}}  \right)^{5} \left( \frac{\sqrt{F}}{100 \, {\rm TeV}} \right)^4
\eea
where $F$ is the $F$-term component of the spurion superfield associated with the Goldstino direction. For the case 
$\,m_{\rm NLSP}  \sim 200$ GeV and assuming that $\Lambda \sim 50$ TeV we get
\bea
&&L \sim 0.008 \,\,\mbox{cm} \,\,\, \mbox{for\, models\, with\, low-scale\, gauge\, mediation,}\,\, M = 10^5 \,\,\mbox{GeV} \nonumber\\
&&L \sim 0.8 \,\,\mbox{m} \,\,\, \mbox{for\, models\, with\, intermediate-scale\, gauge\, mediation,}\,\, M = 10^8 \,\, \mbox{GeV} \nonumber\\
&&L \sim 78000 \,\,\mbox{km} \,\,\, \mbox{for\, models\, with\, high-scale\, gauge\, mediation,}\,\, M = 10^{12} \,\, \mbox{GeV} \nonumber
\eea
Hence, depending on the scale of supersymmetry breaking $\sqrt{F}$, the NLSP can be short- or long-lived. For the models considered in this chapter the NLSP always 
decays in the detector. As far as the LHC is concerned the most relevant decay chains are those involving 
the gluino and the squarks. If the gluino is heavier than the squarks ($m_{\tilde{g}} > m_{\tilde{q}}$), which is always the case for the points in parameter space we have
chosen, the two-body decay of the gluino into a squark and an antiquark $\tilde{g} \to \tilde{q} \bar{q}$ is kinematically possible. The squark then decays further into 
a quark and a chargino or neutralino. In the Tables \ref{tab8} and \ref{tab9} we have summarized the most relevant decay chains for gluinos and squarks for parameter point 4.

\begin{table}[h]\label{gluinojet}
\caption{Decays for the gluino, total width $\Gamma = 115.6$ GeV\label{tab8}}
\begin{center}
\begin{tabular}{|c|c|c|c|c|c|c|c|c|}
\hline
\quad  $\tilde{g} \to \,$  &   jj    & jj + l &  jj + 2l &  jj + 3l &   jjjj   &  jjjj + l    \\
\hline
\quad   Percentage         & 16.72 \% &   71.55 \% &  2.48 \%    &  0.48 \%      & 0.59 \% &    8.16 \%  \\
\hline
\end{tabular}
\end{center}
\end{table}

\begin{table}[h]\label{squarkjet}
\caption{Decays for the squarks, total width $\Gamma = 79.1$ GeV\label{tab9}}
\begin{center}
\begin{tabular}{|c|c|c|c|c|c|c|c|c|}
\hline
\quad  $\tilde{q} \to \,$  &   j    & j+l & j+2l &  j+3l & jjj   &  jjj+l    \\
\hline
\quad   Percentage         & 18.21 \% &   76.56 \% &  1.06 \%    &  0.24 \%      & 0.26 \% &    3.67 \%  \\
\hline
\end{tabular}
\end{center}
\end{table}

As of spring 2012, the strongest experimental lower bounds on the masses of superpartners come 
from searches for events with jets and missing transverse momentum at the ATLAS detector \cite{confnote1,confnote2,Aad:2011ib}. 
This signature typically arises in cascade decays of squarks and gluinos from strong pair production 
$$pp \longrightarrow \tilde{g}\tilde{g} , \tilde{g}\tilde{q}, \tilde{q}\tilde{q}\,. $$
Limits are given in terms of MSUGRA/CMSSM parameters and within a simplified
squark-gluino-neutralino model. The latter does not include gluino decays to
third generation squarks which results in an enhanced branching ratio to first
and second generation quarks. Furthermore, the neutralino is assumed massless,
giving maximal phase space.  These two simplifications lead to an enhanced
exclusion range compared to more generic models, with $m_{\tilde{q}}\gtrsim
1.8$ TeV for $m_{\tilde{g}}\sim 1$ TeV and $m_{\tilde{q}}\gtrsim 1.3$ TeV for
$m_{\tilde{g}}\gtrsim 2$ TeV \cite{confnote1}.  However, these large exclusion
bounds must be taken with a grain of salt if they are to be applied to any
particular SUSY model. In the MSUGRA model for example, the exclusion limit for
the average squark mass is reduced to $m_{\tilde{q}}> 1.3$ TeV for
$m_{\tilde{g}}\sim 1$ TeV. The limits might be reduced even further for the model studied in this
work, where decays to jets and missing transverse momentum are not the dominant decay mode of
the strongly pair-produced superpartners. The situation is illustrated in
Tables \ref{tab8} and \ref{tab9} for parameter point 4. Less than $20\%$ of squarks and gluinos
decay to jets and an invisible NLSP, giving us less than $5\%$ of the
$\tilde{q}\tilde{q}, \tilde{q} \tilde{g}, \tilde{g}\tilde{g}$ pairs decaying
purely to jets and missing transverse momentum.  This is due to the fact that
in our NMSSM-like model, the heavy neutralinos and charginos which
predominantly occur in squark and gluino decays, generically decay to leptons
and sleptons.  For the parameter point 4 this means that
roughly 3/4 of decays are to jets, one or more leptons, and missing energy.
Consequently, roughly $80\%$ of pair produced superpartners yield a final state
with jets, at least one lepton, and missing energy. There is a number of
relevant LHC searches for this type of signature \cite{confnote3,
atlas2,atlas3,atlas5,Aad:2011cwa}.  One can get a rough estimate of the
expected exclusion power. In the case of MSUGRA/CMSSM, the search for jets, one
isolated lepton and missing energy gives roughly the
same bounds as the 0-lepton channel for gluino masses $m_{\tilde{g}}<1.2$ TeV, but
considerably weaker constraints on $m_{\tilde q}$ for $m_{\tilde g}>1.2$ TeV (see for example
Figure 10 in \cite{confnote3}). However, we expect an enhancement of the exclusion from
n-lepton final states in our model due to the suppression of 0-lepton decays. This requires 
a careful analysis of the
various final states with leptons in the context of lopsided gauge mediation in
the NMSSM, and is beyond the scope of this work. Furthermore, there are bounds on 
weak production of electroweak gauginos decaying to leptons \cite{atlas1}. 

If one takes the hints\cite{higgsatlas, higgscms} towards a $\,\,\sim 125$ GeV Higgs boson at the LHC at face
value \cite{Klute:2012pu}, there is a mild tension between the $\gamma \gamma$
excess and parameter points 6 and 8 which feature a lightest pseudoscalar $A_1$
with $2m_{A_1}< m_{h^0}$. In those cases, this decay mode dominates and
$B(h^0\rightarrow A_1 A_1)\approx 30\%\dots 40\%$ with $B(A_1\rightarrow
b\overline b)\approx 90\%$, and thus all other branching fractions are reduced
accordingly. At the same time, this marginally improves the situation for the
$WW$ and $ZZ$ final state.

\section{Conclusions}

In the present paper we investigated a version of the Next to Minimal Supersymmetric Standard Model
in which the soft masses in the Higgs sector obey the non-trivial hierarchy $\,m_{H_d}^2 \gg m_{H_u}^2$ at the weak scale. 
As argued in Chapter 5 this relation can be obtained by a suitable embedding of the low-energy model into a gauge-mediated 
scenario. Our investigation is prompted by a well-known result due to H. Murayama, A. de Gouvea and A. Friedland 
\cite{deGouvea:1997cx} which states that the $\mathbb{Z}_3$-invariant NMSSM is incompatible with the minimal version 
of gauge-mediated supersymmetry breaking.

We described three important effects which occur within the framework of our model.
The first effect is related to the size of the VEV of the gauge singlet superfield $N$. Recall that in the MGM version of the NMSSM 
the small value of $\,s = \langle N \rangle\,$ is the very reason why the model fails to produce phenomenologically
viable spectra. The lopsided extension of this scenario solves that problem by utilizing the non-standard running
of the gauge singlet soft mass $m_N^2$. To make this statement more precise we note that the one-loop $\beta$-function of $m_N^2$
is comparatively large due to the presence of the dominant $m_{H_d}^2$ term. As a consequence the $m_N^2$ mass is shifted 
towards large negative values along its RG trajectory which eventually triggers a sizeable VEV for the scalar component of $N$.

The second effect is related to the mass $m_{h^0}$ of the lightest CP-even Higgs boson. We showed that within the lopsided NMSSM one
can easily generate large soft trilinear couplings at the messenger scale and therefore increase the one-loop radiative correction 
associated with top/stop loops, leading to a Higgs mass in the range of 125 GeV.

The third effect is the inverted mass hierarchy in the slepton sector. As we argued this effect can be traced back to the atypical 
RG running of the soft masses of the left-handed and right-handed sleptons. To be more precise, the $\,m_{H_d}^2\,$ term enhances 
the $\,\footnotesize{\frac{6}{5}}\,Y_{\tilde{f}}\,g_1^2 \, {\rm Tr}\,[\,Y_{\tilde{f}} \, m_{\tilde{f}}^2\,]\,$ contribution to the 
one-loop $\beta$-function of the slepton masses leading to the relation $\,m_{\tilde{L}_R}^2 > m_{\tilde{L}_L}^2\,$ at the electroweak scale. Something similar 
happens in the squark sector although the effect is much less prominent since the RG running of the colour charged sparticles is dominated by 
the gluino contribution.

As we showed in section 6.3 the UV completion of the lopsided NMSSM can look quite complicated. In particular we had to rely on
three completely different sets of operators in order to induce the necessary extra contributions to the mass terms in the 
Higgs sector. It would be interesting to investigate whether a simpler realization of the lopsided NMSSM is possible.

\section*{Acknowledgements}

ID would like to thank Tilman Plehn for the financial support during the final stages of this work.
Both authors would like to thank Tilman Plehn and Arthur Hebecker for useful discussions and suggestions.

\begin{appendix}
\section{One-loop RGEs for the soft squark, slepton and Higgs masses in the NMSSM}

In this appendix we list the one-loop RGEs for the soft mass terms in the NMSSM in the presence of a tree-level $\mu$-term (see \cite{Ellwanger:2009dp}
and \cite{King:1995vk}). This is the low-energy model which is obtained after integrating out all messenger superfields in eq.(\ref{fullUV}) 
at their respective decoupling scales. It will be useful to introduce the following quantities:
\bea
&&X_t = m_{Q_3}^2 + m_{U_3}^2 + m_{H_u}^2 + A_t^2 \\
\noalign{\smallskip}
&&X_b = m_{Q_3}^2 + m_{D_3}^2 + m_{H_d}^2 + A_b^2 \\
\noalign{\smallskip}
&&X_{\tau} = m_{L_3}^2 + m_{E_3}^2 + m_{H_d}^2 + A_{\tau}^2 \\
\noalign{\smallskip}
&&X_{\lambda} = m_{H_u}^2 + m_{H_d}^2 + m_{N}^2 + A_{\lambda}^2 \\
\noalign{\smallskip}
&&X_{\kappa} = 3 m_{N}^2 +  A_{\kappa}^2 \\
\noalign{\smallskip}
&&S \equiv {\rm Tr}[Y_{\tilde{f}} \, m_{\tilde{f}}^2] = m_{H_u}^2 - m_{H_d}^2 + 
{\rm Tr}[\mathbf{m_Q^2 - m_L^2 - 2 m_{U}^2 + m_{D}^2 + m_{E}^2}\,]
\eea
where the boldface $\,\mathbf{m}$'s are mass matrices in family space and $\,Y_{\tilde{f}}\,$ 
denotes the hypercharge of the sfermion field $\tilde{f}$.
\bea
&&16\pi^2 \, \frac{d}{dt} \,m_{\tilde{Q}_i}^2 = 2 \delta_{i3}\,y_t^2 X_t \,+ \, 2 \delta_{i3}\,y_b^2 X_b \,
- \,\frac{32}{3} g_3^2 \,|M_3|^2 \, - \,6 g_2^2 \,|M_2|^2 \,- \frac{2}{15} g_1^2 \,|M_1|^2 \, + \, \frac{1}{3}\, g_1^2 S \nonumber \\
&&16\pi^2 \, \frac{d}{dt} \,m_{\tilde{\bar{u}}_i}^2  = 2 \delta_{i3}\,y_t^2 X_t \, 
- \,\frac{32}{3} g_3^2 \,|M_3|^2 \, - \, \frac{32}{15} g_1^2 \,|M_1|^2 \, - \, \frac{4}{3}\, g_1^2 S \nonumber \\
&&16\pi^2 \, \frac{d}{dt} \,m_{\tilde{\bar{d}}_i}^2  = 2 \delta_{i3}\,y_b^2 X_b  - \,\frac{32}{3} g_3^2 \,|M_3|^2 \, - \,\frac{8}{15} g_1^2 \,|M_1|^2 \, + \, \frac{2}{3}\, g_1^2 S  \nonumber\\ 
&&16\pi^2 \, \frac{d}{dt} \,m_{\tilde{L}_i}^2 = 2 \delta_{i3}\,y_{\tau}^2 X_{\tau}\, - \,6 g_2^2 \,|M_2|^2 \,- \frac{6}{5} g_1^2 \,|M_1|^2 \, - \, g_1^2 S \nonumber \\
&&16\pi^2 \, \frac{d}{dt} \,m_{\tilde{e}_i}^2  = 2 \delta_{i3}\,y_{\tau}^2 X_{\tau}\, - \,\frac{24}{5} g_1^2 \,|M_1|^2 \, + \, 2\, g_1^2 S \nonumber \\
&&16\pi^2 \, \frac{d}{dt} \,m_{H_u}^2 = 2 \,y_t^2 X_t \,+ \, 2 \,\lambda^2 X_{\lambda} \,
- \,6 g_2^2 \,|M_2|^2 \,- 2 g_1^2 \,|M_1|^2 \, + \, g_1^2 S \nonumber \\
&&16\pi^2 \, \frac{d}{dt} \,m_{H_d}^2 = 6 \,y_b^2 X_b \,+ \, 2\,y_{\tau}^2 X_{\tau} \,+ \, 
2 \,\lambda^2 X_{\lambda}\, - \,6 g_2^2 \,|M_2|^2 \,- 2 g_1^2 \,|M_1|^2 \, - \, g_1^2 S \nonumber \\
&&16\pi^2 \, \frac{d}{dt} \,m_{N}^2 = 2 \,\lambda^2 X_{\lambda} \,+ \, 2 \,\kappa^2 X_{\kappa}
\eea

\section{Coefficient functions for the induced mass terms in the Higgs sector}

In this appendix we list the coefficient functions appearing in the formula for the induced mass terms in
the Higgs sector:
\bea
P(x,y)  &=&  \frac{x^2 (1-y)^2}{(x^2-1)^3}\, \big(\,2(1-x^2) \,+\, (1+x^2)\log x^2\,\big) \nonumber\\
Q(x,y)  &=&  \frac{x}{(x^2-1)^2}\, \big(\,(x^2-1)(1-y) \,+\, (y-x^2)\log x^2\,\big) \nonumber\\
R(x,y)  &=&  \frac{x}{(x^2-1)^3}\, \big(\,(1-x^4)(1-y)^2 \,+\, [\,2x^2(1+y^2) - y(1+x^2)^2\,]\log x^2\,\big) \nonumber\\
S(x,y)  &=&  \frac{1}{(x^2-1)^2}\, \big(\,(x^2-1)(1-x^2y) \,-\, x^2 (1-y) \log x^2\,\big) 
\eea
All four functions have the property $\,|P(x,y)|$,  $|Q(x,y)|$,  $|R(x,y)|$, $|S(x,y)| \le 1$. Additionally 
we have $P(x,y=1) = 0\,$ and $\,Q(x,y=1) = R(x,y=1)$.

\end{appendix}


\newpage



\begin{thebibliography}{99}
 








%\cite{Giudice:1998bp}
\bibitem{Giudice:1998bp}
  G.~F.~Giudice and R.~Rattazzi,
  ``Theories with gauge mediated supersymmetry breaking,''
  Phys.\ Rept.\  {\bf 322} (1999) 419
  [hep-ph/9801271].
  %%CITATION = HEP-PH/9801271;%%

%\cite{Gaume:1982cw}
\bibitem{Gaume:1982cw}
      L.~Alvarez-Gaume, M.~Claudson, M.~Wise,
      Nucl. Phys. {\bf B 207} (1982) 96;\\
%\bibitem{Dine:1982df}
     M.~Dine and W.~Fischler, 
     Phys. Lett. {\bf B 110} (1982) 227;\\
%\cite{Dine:1993yw}
     S.~Dimopoulos, M.~Dine, S.~Raby, T.~Scott
     Phys. Rev. Lett. {\bf 76} (1996) 3494-3497 .


\bibitem{Dine:1993yw}
  M.~Dine and A.~E.~Nelson,
  ``Dynamical supersymmetry breaking at low-energies,''
  Phys.\ Rev.\ D {\bf 48} (1993) 1277
  [hep-ph/9303230];\\
  %%CITATION = HEP-PH/9303230;%%
%\cite{Dine:1994vc}
%\bibitem{Dine:1994vc}
  M.~Dine, A.~E.~Nelson and Y.~Shirman,
  ``Low-energy dynamical supersymmetry breaking simplified,''
  Phys.\ Rev.\ D {\bf 51} (1995) 1362
  [hep-ph/9408384];\\
  %%CITATION = HEP-PH/9408384;%%
%\cite{Dine:1995ag}
%\bibitem{Dine:1995ag}
  M.~Dine, A.~E.~Nelson, Y.~Nir and Y.~Shirman,
  ``New tools for low-energy dynamical supersymmetry breaking,''
  Phys.\ Rev.\ D {\bf 53} (1996) 2658
  [hep-ph/9507378].
  %%CITATION = HEP-PH/9507378;%%




%\cite{Ellis:1981ts}
\bibitem{Ellis:1981ts}
  J.~R.~Ellis and D.~V.~Nanopoulos,
  ``Flavor Changing Neutral Interactions In Broken Supersymmetric Theories,''
  Phys.\ Lett.\  B {\bf 110} (1982) 44.
  %%CITATION = PHLTA,B110,44;%%



%\cite{Peccei:1977hh}
\bibitem{Peccei:1977hh}
  R.~D.~Peccei and H.~R.~Quinn,
  ``CP Conservation In The Presence Of Instantons,''
  Phys.\ Rev.\ Lett.\  {\bf 38} (1977) 1440.
  %%CITATION = PRLTA,38,1440;%%

%\cite{Peccei:1977ur}
\bibitem{Peccei:1977ur}
  R.~D.~Peccei and H.~R.~Quinn,
  ``Constraints Imposed By CP Conservation In The Presence Of Instantons,''
  Phys.\ Rev.\  D {\bf 16} (1977) 1791.
  %%CITATION = PHRVA,D16,1791;%%



%\cite{Guidice:1988gm}
\bibitem{Guidice:1988gm}
  G.~F.~Giudice and A.~Masiero,
  ``A natural solution to the $\mu$-problem in supergravity theories,''
  Phys.\ Lett.\ B {\bf 206} (1988) .
  %%CITATION = HEP-PH/0000000;%%




%\cite{Dvali:1996cu}
\bibitem{Dvali:1996cu}
  G.~R.~Dvali, G.~F.~Giudice and A.~Pomarol,
  ``The $\mu$-Problem in Theories with Gauge-Mediated Supersymmetry Breaking,''
  Nucl.\ Phys.\  B {\bf 478} (1996) 31
  [arXiv:hep-ph/9603238].
  %%CITATION = NUPHA,B478,31;%%


%\cite{Csaki:2008sr}
\bibitem{Csaki:2008sr}
  C.~Csaki, A.~Falkowski, Y.~Nomura and T.~Volansky,
  ``New Approach to the mu-Bmu Problem of Gauge-Mediated Supersymmetry
  Breaking,''
  Phys.\ Rev.\ Lett.\  {\bf 102} (2009) 111801
  [arXiv:0809.4492 [hep-ph]].
  %%CITATION = PRLTA,102,111801;%%



%\cite{DeSimone:2011va}
\bibitem{DeSimone:2011va}
  A.~De Simone, R.~Franceschini, G.~F.~Giudice, D.~Pappadopulo and R.~Rattazzi,
  ``Lopsided Gauge Mediation,''
  JHEP {\bf 1105} (2011) 112
  [arXiv:1103.6033 [hep-ph]].
  %%CITATION = JHEPA,1105,112;%%



%\cite{Ellis:1989pr}
\bibitem{Ellis:1989pr}
 J.~Ellis, F.~Gunion, H.~E.~Haber, L.~Roszkowski and F.~Zwirner,
 ``Higgs bosons in a nonminimal supersymmetric model,''
 Phys.\ Rev.\ D. {\bf 39. 844} (1989) 
 

%\cite{Drees:1988fc}
\bibitem{Drees:1988fc}
  M.~Drees,
  ``Supersymmetric Models with Extended Higgs Sector,''
  Int.\ J.\ Mod.\ Phys.\  A {\bf 4} (1989) 3635.
  %%CITATION = IMPAE,A4,3635;%%


%\cite{Ellwanger:2004xm}
\bibitem{Ellwanger:2004xm} 
  U.~Ellwanger, J.~F.~Gunion and C.~Hugonie,
  ``NMHDECAY: A Fortran code for the Higgs masses, couplings and decay widths in the NMSSM,''
  JHEP {\bf 0502}, 066 (2005)
  [hep-ph/0406215].
  %%CITATION = HEP-PH/0406215;%%
%\cite{Ellwanger:2006rn}
\bibitem{Ellwanger:2006rn}
  U.~Ellwanger and C.~Hugonie,
  ``NMSPEC: A Fortran code for the sparticle and Higgs masses in the NMSSM with GUT scale boundary conditions,''
  Comput.\ Phys.\ Commun.\  {\bf 177} (2007) 399
  [hep-ph/0612134].
  %%CITATION = HEP-PH/0612134;%%

%\cite{Franke:1995tc}
\bibitem{Franke:1995tc} 
  F.~Franke and H.~Fraas,
  ``Neutralinos and Higgs bosons in the next-to-minimal supersymmetric standard model,''
  Int.\ J.\ Mod.\ Phys.\ A {\bf 12}, 479 (1997)
  [hep-ph/9512366].
  %%CITATION = HEP-PH/9512366;%%


%\cite{Ellwanger:1993xa}
\bibitem{Ellwanger:1993xa} 
  U.~Ellwanger, M.~Rausch de Traubenberg and C.~A.~Savoy,
  ``Particle spectrum in supersymmetric models with a gauge singlet,''
  Phys.\ Lett.\ B {\bf 315}, 331 (1993)
  [hep-ph/9307322].
  %%CITATION = HEP-PH/9307322;%%


%\cite{BasteroGil:2000bw}
\bibitem{BasteroGil:2000bw} 
  M.~Bastero-Gil, C.~Hugonie, S.~F.~King, D.~P.~Roy and S.~Vempati,
  ``Does LEP prefer the NMSSM?,''
  Phys.\ Lett.\ B {\bf 489}, 359 (2000)
  [hep-ph/0006198].
  %%CITATION = HEP-PH/0006198;%%

%\cite{King:1995vk}
\bibitem{King:1995vk} 
  S.~F.~King and P.~L.~White,
  ``Resolving the constrained minimal and next-to-minimal supersymmetric standard models,''
  Phys.\ Rev.\ D {\bf 52}, 4183 (1995)
  [hep-ph/9505326]. \textit{cite for $\beta$-function}
  %%CITATION = HEP-PH/9505326;%%


%\cite{Abel:1995wk}
\bibitem{Abel:1995wk} 
  S.~A.~Abel, S.~Sarkar and P.~L.~White,
  ``On the cosmological domain wall problem for the minimally extended supersymmetric standard model,''
  Nucl.\ Phys.\ B {\bf 454}, 663 (1995)
  [hep-ph/9506359].
  %%CITATION = HEP-PH/9506359;%%

%\cite{Panagiotakopoulos:1998yw}
\bibitem{Panagiotakopoulos:1998yw} 
  C.~Panagiotakopoulos and K.~Tamvakis,
  ``Stabilized NMSSM without domain walls,''
  Phys.\ Lett.\ B {\bf 446}, 224 (1999)
  [hep-ph/9809475].
  %%CITATION = HEP-PH/9809475;%%


%\cite{Derendinger1984307}
\bibitem{Derendinger1984307}
  J.~P.~Derendinger and C.~A.~Savoy
  ``Quantum effects and SU(2)xU(1) breaking in supergravity gauge theories,''
  Nucl.\ Phys.\ B {\bf 237} (1984)

%\cite{Ellwanger:2005fh}
\bibitem{Ellwanger:2005fh} 
  U.~Ellwanger and C.~Hugonie,
  ``Yukawa induced radiative corrections to the lightest Higgs boson mass in the NMSSM,''
  Phys.\ Lett.\ B {\bf 623}, 93 (2005)
  [hep-ph/0504269].
  %%CITATION = HEP-PH/0504269;%%

%\cite{Degrassi:2009yq}
\bibitem{Degrassi:2009yq} 
  G.~Degrassi and P.~Slavich,
  ``On the radiative corrections to the neutral Higgs boson masses in the NMSSM,''
  Nucl.\ Phys.\ B {\bf 825}, 119 (2010)
  [arXiv:0907.4682 [hep-ph]]. \textit{radiative NMSSM specific one-loop corrections}
  %%CITATION = ARXIV:0907.4682;%%

%\cite{Ellwanger:1999ji}
\bibitem{Ellwanger:1999ji} 
  U.~Ellwanger and C.~Hugonie,
  ``Masses and couplings of the lightest Higgs bosons in the (M+1)SSM,''
  Eur.\ Phys.\ J.\ C {\bf 25}, 297 (2002)
  [hep-ph/9909260].
  %%CITATION = HEP-PH/9909260;%%

%\cite{Masip:1998jc}
\bibitem{Masip:1998jc} 
  M.~Masip, R.~Munoz-Tapia and A.~Pomarol,
  ``Limits on the mass of the lightest Higgs in supersymmetric models,''
  Phys.\ Rev.\ D {\bf 57}, R5340 (1998)
  [hep-ph/9801437].
  %%CITATION = HEP-PH/9801437;%%

%\cite{Yeghian:1999kr}
\bibitem{Yeghian:1999kr} 
  G.~K.~Yeghian,
  ``Upper bound on the lightest Higgs mass in supersymmetric theories,''
  hep-ph/9904488.
  %%CITATION = HEP-PH/9904488;%%


%\cite{Katz:2009qx}
\bibitem{Katz:2009qx}
  A.~Katz and B.~Tweedie,
  ``Signals of a Sneutrino (N)LSP at the LHC,''
  Phys.\ Rev.\  D {\bf 81} (2010) 035012
  [arXiv:0911.4132 [hep-ph]].
  %%CITATION = PHRVA,D81,035012;%%


%\cite{Kang:2011az}
\bibitem{Kang:2011az}
  Z.~Kang, T.~Li, T.~Liu and J.~M.~Yang,
  ``The Minimal Solution to the $\mu/B_{\mu}$ Problem in Gauge Mediation,''
  arXiv:1109.4993 [hep-ph].
  %%CITATION = ARXIV:1109.4993;%%





%\cite{Ellwanger:2009dp}
\bibitem{Ellwanger:2009dp}
  U.~Ellwanger, C.~Hugonie and A.~M.~Teixeira,
  ``The Next-to-Minimal Supersymmetric Standard Model,''
  Phys.\ Rept.\  {\bf 496} (2010) 1
  [arXiv:0910.1785 [hep-ph]].
  %%CITATION = PRPLC,496,1;%%

%\cite{Komargodski:2008ax}
\bibitem{Komargodski:2008ax}
  Z.~Komargodski and N.~Seiberg,
  ``mu and General Gauge Mediation,''
  JHEP {\bf 0903} (2009) 072
  [arXiv:0812.3900 [hep-ph]].
  %%CITATION = JHEPA,0903,072;%%


%\cite{Draper:2011aa}
\bibitem{Draper:2011aa} 
  P.~Draper, P.~Meade, M.~Reece and D.~Shih,
  ``Implications of a 125 GeV Higgs for the MSSM and Low-Scale SUSY Breaking,''
  arXiv:1112.3068 [hep-ph].
  %%CITATION = ARXIV:1112.3068;%%


%\cite{deGouvea:1997cx}
\bibitem{deGouvea:1997cx}
  A.~de Gouvea, A.~Friedland and H.~Murayama,
  ``Next-to-minimal supersymmetric standard model with the gauge mediation  of
  supersymmetry breaking,''
  Phys.\ Rev.\  D {\bf 57} (1998) 5676
  [arXiv:hep-ph/9711264].
  %%CITATION = PHRVA,D57,5676;%%



%\cite{Giudice:2006sn}
\bibitem{Giudice:2006sn}
  G.~F.~Giudice and R.~Rattazzi,
  ``Living Dangerously with Low-Energy Supersymmetry,''
  Nucl.\ Phys.\  B {\bf 757} (2006) 19
  [arXiv:hep-ph/0606105].
  %%CITATION = NUPHA,B757,19;%%


%\cite{Babu:2008ge}
\bibitem{Babu:2008ge}
  K.~S.~Babu, I.~Gogoladze, M.~U.~Rehman and Q.~Shafi,
  ``Higgs Boson Mass, Sparticle Spectrum and Little Hierarchy Problem in
  Extended MSSM,''
  Phys.\ Rev.\  D {\bf 78} (2008) 055017
  [arXiv:0807.3055 [hep-ph]].
  %%CITATION = PHRVA,D78,055017;%%





%\cite{Mason:2009iq}
\bibitem{Mason:2009iq}
  J.~D.~Mason,
  ``Gauge Mediation with a small mu term and light squarks,''
  Phys.\ Rev.\  D {\bf 80} (2009) 015026
  [arXiv:0904.4485 [hep-ph]].
  %%CITATION = PHRVA,D80,015026;%%


 %\cite{Dine:2009gy}
\bibitem{Dine:2009gy}
  M.~Dine,
  ``Supersymmetry Breaking at Low Energies,''
  Nucl.\ Phys.\ Proc.\ Suppl.\  {\bf 192-193} (2009) 40
  [arXiv:0901.1713 [hep-ph]].
  %%CITATION = NUPHZ,192-193,40;%%

%\cite{Martin:2009bg}
\bibitem{Martin:2009bg}
  S.~P.~Martin,
  ``Extra vector-like matter and the lightest Higgs scalar boson mass in
  low-energy supersymmetry,''
  Phys.\ Rev.\  D {\bf 81} (2010) 035004
  [arXiv:0910.2732 [hep-ph]].
  %%CITATION = PHRVA,D81,035004;%%




%\cite{Martin:2010dc}
\bibitem{Martin:2010dc}
  S.~P.~Martin,
  ``Raising the Higgs mass with Yukawa couplings for isotriplets in vector-like
  extensions of minimal supersymmetry,''
  Phys.\ Rev.\  D {\bf 82} (2010) 055019
  [arXiv:1006.4186 [hep-ph]].
  %%CITATION = PHRVA,D82,055019;%%

%\cite{Barbieri:2006bg}
\bibitem{Barbieri:2006bg}
  R.~Barbieri, L.~J.~Hall, Y.~Nomura and V.~S.~Rychkov,
  ``Supersymmetry without a Light Higgs Boson,''
  Phys.\ Rev.\  D {\bf 75} (2007) 035007
  [arXiv:hep-ph/0607332].
  %%CITATION = PHRVA,D75,035007;%%


%\cite{Barbieri:2007tu}
\bibitem{Barbieri:2007tu}
  R.~Barbieri, L.~J.~Hall, A.~Y.~Papaioannou, D.~Pappadopulo and V.~S.~Rychkov,
  ``An alternative NMSSM phenomenology with manifest perturbative
  unification,''
  JHEP {\bf 0803} (2008) 005
  [arXiv:0712.2903 [hep-ph]].
  %%CITATION = JHEPA,0803,005;%%





\bibitem{atlas1} 
  G.~Aad {\it et al.}  [ATLAS Collaboration],
  ``Search for supersymmetry in events with three leptons and missing transverse momentum in sqrt(s) = 7 TeV pp collisions with the ATLAS detector,''
  arXiv:1204.5638 [hep-ex].
  %%CITATION = ARXIV:1204.5638;%%
%\cite{}
\bibitem{atlas2} 
  G.~Aad {\it et al.}  [ATLAS Collaboration],
  ``Search for supersymmetry with jets, missing transverse momentum and at least one hadronically decaying tau lepton in proton-proton collisions at sqrt(s) = 7 TeV with the ATLAS detector,''
  arXiv:1204.3852 [hep-ex].
  %%CITATION = ARXIV:1204.3852;%%
%\cite{}
\bibitem{atlas3} 
  G.~Aad {\it et al.}  [ATLAS Collaboration],
  ``Search for events with large missing transverse momentum, jets, and at least two tau leptons in 7 TeV proton-proton collision data with the ATLAS detector,''
  arXiv:1203.6580 [hep-ex].
  %%CITATION = ARXIV:1203.6580;%%
%\cite{}
\bibitem{atlas4} 
  G.~Aad {\it et al.}  [ATLAS Collaboration],
  ``Search for supersymmetry in pp collisions at sqrt(s) = 7 TeV in final states with missing transverse momentum and b-jets with the ATLAS detector,''
  arXiv:1203.6193 [hep-ex].
  %%CITATION = ARXIV:1203.6193;%%
%\cite{}
\bibitem{atlas5} 
  G.~Aad {\it et al.}  [ATLAS Collaboration],
  ``Search for gluinos in events with two same-sign leptons, jets and missing transverse momentum with the ATLAS detector in pp collisions at sqrt(s) = 7 TeV,''
  arXiv:1203.5763 [hep-ex].
  %%CITATION = ARXIV:1203.5763;%%
%\cite{Aad:2011cw}
\bibitem{Aad:2011cw} 
  G.~Aad {\it et al.}  [ATLAS Collaboration],
  ``Search for scalar bottom pair production with the ATLAS detector in pp Collisions at sqrt{s} = 7 TeV,''
  arXiv:1112.3832 [hep-ex].
  %%CITATION = ARXIV:1112.3832;%%
%\cite{Aad:2011cwa}
\bibitem{Aad:2011cwa} 
  G.~Aad {\it et al.}  [ATLAS Collaboration],
  ``Searches for supersymmetry with the ATLAS detector using final states with two leptons and missing transverse momentum in sqrt{s} = 7 TeV proton-proton collisions,''
  Phys.\ Lett.\ B {\bf 709}, 137 (2012)
  [arXiv:1110.6189 [hep-ex]].
  %%CITATION = ARXIV:1110.6189;%%
%\cite{Aad:2011yh}
\bibitem{Aad:2011yh} 
  G.~Aad {\it et al.}  [ATLAS Collaboration],
  ``Search for Massive Colored Scalars in Four-Jet Final States in sqrt{s}=7 TeV proton-proton collisions with the ATLAS Detector,''
  Eur.\ Phys.\ J.\ C {\bf 71}, 1828 (2011)
  [arXiv:1110.2693 [hep-ex]].
  %%CITATION = ARXIV:1110.2693;%%
%\cite{Aad:2011ib}
\bibitem{Aad:2011ib} 
  G.~Aad {\it et al.}  [ATLAS Collaboration],
  ``Search for squarks and gluinos using final states with jets and missing transverse momentum with the ATLAS detector in sqrt(s) = 7 TeV proton-proton collisions,''
  Phys.\ Lett.\ B {\bf 710}, 67 (2012)
  [arXiv:1109.6572 [hep-ex]].
  %%CITATION = ARXIV:1109.6572;%%
\bibitem{confnote1}
  [ATLAS Collaboration],
 ``Search for squarks and gluinos with the ATLAS detector using final states
 with jets and missing transverse momentum and 4.7 $fb^{???1}$ of $\sqrt{s}$ = 7 TeV
proton-proton collision data
 ''
ATLAS-CONF-2012-033
March 11, 2012

\bibitem{confnote2}
  [ATLAS Collaboration],
 ``Hunt for new phenomena using large jet multiplicities and missing
transverse momentum with ATLAS in L = 4.7 $fb^{???1}$ of $\sqrt{s}$ = 7 TeV
proton-proton collisions
 ''
ATLAS-CONF-2012-037
Match 11, 2012
\bibitem{confnote3}
  [ATLAS Collaboration],
 ``Further search for supersymmetry at $\sqrt{s}$ = 7 TeV in final states with jets,
missing transverse momentum and one isolated lepton
 ''
ATLAS-CONF-2012-041
March 16, 2012


\bibitem{higgsatlas}
The ATLAS Colloboration,
%``Search for the Standard Model Higgs boson in the decay channel H->ZZ(*)->4l with 4.8 fb-1 of pp collision data at sqrt(s) = 7 TeV with ATLAS,''
  Phys.\ Lett.\ B {\bf 710}, 383 (2012);
  %[arXiv:1202.1415 [hep-ex]];
  %%CITATION = ARXIV:1202.1415;%%
%``Search for the Standard Model Higgs boson in the diphoton decay channel with 4.9 fb-1 of pp collisions at sqrt(s)=7 TeV with ATLAS,''
  Phys.\ Rev.\ Lett.\  {\bf 108}, 111803 (2012);
  %[arXiv:1202.1414 [hep-ex]];
  %%CITATION = ARXIV:1202.1414;%%
%``Combined search for the Standard Model Higgs boson using up to 4.9 fb-1 of pp collision data at sqrt(s) = 7 TeV with the ATLAS detector at the LHC,''
  Phys.\ Lett.\ B {\bf 710}, 49 (2012);
  %[arXiv:1202.1408 [hep-ex]];
  %%CITATION = ARXIV:1202.1408;%%
  %``Search for the Higgs boson in the H->WW(*)->lvlv decay channel in pp collisions at sqrt{s} = 7 TeV with the ATLAS detector,''
  Phys.\ Rev.\ Lett.\  {\bf 108}, 111802 (2012);
  %[arXiv:1112.2577 [hep-ex]];
  %%CITATION = ARXIV:1112.2577;%%
  %ATLAS-CONF-2012-012; ATLAS-CONF-2012-014; ATLAS-CONF-2012-015.

\bibitem{higgscms}
The CMS Collaboration,
%``Search for the standard model Higgs boson decaying to W+W- in the fully leptonic final state in pp collisions at sqrt(s) = 7 TeV",
Phys.\ Lett.\ B {\bf 710}, 91 (2012);
%arXiv:1202.1489[hep-ex];
%The CMS Collaboration,
%``Search for the standard model Higgs boson in the decay channel H to ZZ to 4l in pp collisions at sqrt(s) = 7 TeV",
Phys.\ Rev.\ Lett.\  {\bf 108}, 111804 (2012);
%The CMS Collaboration,
%``Search for the standard model Higgs boson decaying to bottom quarks
%in pp collisions at  sqrt s=7 TeV''
Phys.\ Lett.\ B {\bf 710}, 284 (2012);
%The CMS Collaboration,
%``Search for neutral Higgs bosons decaying to tau pairs in pp collisions at sqrt(s)=7 TeV."
 arXiv:1202.4083[hep-ex];
%The CMS Collaboration,
%``A search using multivariate techniques for a standard model Higgs boson decaying into two photons",
CMS-PAS-HIG-12-001;
%The CMS Collaboration,
%``Search for Neutral Higgs Bosons Decaying into Tau Leptons in the Dimuon Channel with CMS in pp Collisions at 7 TeV",
CMS-PAS-HIG-12-007.



%\cite{Klute:2012pu}
\bibitem{Klute:2012pu} 
  M.~Klute, R.~Lafaye, T.~Plehn, M.~Rauch and D.~Zerwas,
  ``Measuring Higgs Couplings from LHC Data,''
  arXiv:1205.2699 [hep-ph].
  %%CITATION = ARXIV:1205.2699;%%










\end{thebibliography}
\end{document}